\documentclass[pra,twocolumn,preprintnumbers,amsmath,amssymb,nofootinbib]{revtex4}
\usepackage{bbm,bm,slashed}
\usepackage{graphicx,graphics,psfrag,epsfig}
\usepackage{color,array,tabularx}
\usepackage{hyperref}

\definecolor{holger}{rgb}{0.9,0,0}

\definecolor{michael}{rgb}{0.9,0,0.9}
\newcommand{\commentM}[1]{\textbf{\textcolor{michael}{(M: #1)}}}
\definecolor{stefan}{rgb}{0,0.6,0}

\definecolor{lukas}{rgb}{0,0,0.9}

\newcommand{\dslash}{\slashed{\partial}}
\newcommand{\ir}{\mathrm i}
\newcommand{\psiL}{\psi_\mathrm L}
\newcommand{\psiR}{\psi_\mathrm R}
\newcommand{\bpsiL}{\bar\psi_\mathrm L}
\newcommand{\bpsiR}{\bar\psi_\mathrm R}
\newcommand{\Nf}{N_\mathrm{f}}
\newcommand{\NL}{N_\mathrm{L}}

\DeclareMathOperator{\er}{e}
\newcommand{\Eqref}[1]{Eq.~\eqref{#1}}
\newcommand{\pat}{\partial_t}

\begin{document}

\title{Phase transition and critical behavior of d=3 chiral fermion models
  with left/right asymmetry}
\date{\today}
\author{Holger Gies, Lukas Janssen, Stefan
 Rechenberger and Michael M. Scherer}


\affiliation{\mbox{\it Theoretisch-Physikalisches Institut,
 Friedrich-Schiller-Universit{\"a}t Jena,}
\mbox{\it Max-Wien-Platz 1, D-07743 Jena, Germany}
\mbox{\it E-mail: {holger.gies@uni-jena.de, lukas.janssen@uni-jena.de}}
\mbox{\it {stefan.rechenberger@uni-jena.de, michael.scherer@uni-jena.de,}}
}

\begin{abstract} 
We investigate the critical behavior of three-dimensional relativistic fermion
models with a U($\NL$)$_\mathrm{L} \times \mathrm U(1)_{\mathrm{R}}$ chiral
symmetry reminiscent of the Higgs-Yukawa sector of the standard model of
particle physics. We classify all possible four-fermion interaction terms and
the corresponding discrete symmetries. For sufficiently strong correlations in
a scalar parity-conserving channel, the system can undergo a second-order
phase transition to a chiral-symmetry broken phase which is a 3d analog of the
electroweak phase transition. We determine the critical behavior of this phase
transition in terms of the critical exponent $\nu$ and the fermion and scalar
anomalous dimensions for $\NL\geq1$. 
Our models define new universality classes that can serve as prototypes for
studies of strongly correlated chiral fermions. 
\end{abstract}

\maketitle

\section{Introduction}

The interplay between statistical physics and particle physics has stimulated
substantial progress in both areas in the last decades. It has given rise to
the renormalization group (RG)
\cite{Stueckelberg:1953dz,Gell-Mann:fq,Bogolyubov:1956gh} which has led
to a deep understanding of critical phenomena on the statistical-physics side
and of the possible structure of fundamental interactions on the
particle-physics side. In most cases, this interplay arises from analogies on
the technical level of dealing with fluctuations either of statistical or
quantum nature.

In the present work, we intend to push this interplay one step further: we aim
at a construction of a statistical-physics model (or a field-theoretic variant
thereof) which has structural similarities to a crucial building block of the
standard model of particle physics: the Higgs-Yukawa sector. We are motivated
by the fact that a profound understanding of electroweak symmetry breaking in
the standard model or alternative scenarios may require a quantitative control
of fluctuating chiral fermions and bosons beyond perturbation
theory.\footnote{This work has partly been inspired by a recently discovered
  asymptotic-safety scenario for the Higgs-Yukawa sector of the standard model
  which aims at a solution of the triviality and an improvement of the
  hierarchy problem of the standard-model Higgs sector
  \cite{Gies:2009hq,Gies:2009sv}.} This is a challenging problem which is even
lacking a simpler benchmark example where nonperturbative techniques can prove
their validity.

A stage for quantitative comparisons between field-theoretical tools is set by
critical phenomena of lower-dimensional systems, most prominently the
computation of critical exponents of second-order phase transitions. In fact,
relativistic fermionic models such as the Gross-Neveu model in 3d are known to
exhibit a second-order phase transition; corresponding studies of the critical
behavior have been performed by various methods
\cite{HandsEtAl,Rosa:2000ju}. Near the phase transition, the critical behavior
is determined by fluctuations of fermions as well as bosonic bi-fermion
composites; the expectation value of the latter also serves as an order
parameter for the phase transition.

In the same spirit, we devote this work to a construction of chiral
four-fermion models with a U($\NL$)$_\mathrm{L} \times \mathrm
U(1)_{\mathrm{R}}$ symmetry. Such models support a left/right asymmetry
similar to the chiral structure of the Higgs-Yukawa sector. If such models
undergo a phase transition to an ordered phase with broken chiral symmetry
this transition can be viewed as an analogue of the electroweak phase
transition in 3d. The corresponding critical behavior defines new universality
classes, the properties of which are a pure prediction of the system. 

In the past, $d=3$ dimensional fermionic systems with left/right symmetric
chiral symmetries such as QED${}_3$ or the Thirring model have been under
investigation in a variety of scenarios
\cite{Pisarski:1984dj,Appelquist:1985vf,Gomes:1990ed,Hands:2002dv,%
  Maris:1996zg,Mavromatos:2003ss} with applications to condensed-matter
physics, high-$T_c$ cuprate superconductors \cite{Herbut:2002yq} and,
recently, graphene \cite{Herbut:2009qb}.  In some of these models, the number
of fermion flavors serves as a control parameter for a quantum phase
transition. As the critical number of fermions is an important quantity,
nonperturbative information about these models for varying flavor number $\Nf$
is required.  Since good chiral fermion properties for arbitrary $\Nf$ for
massless fermions still represents a challenge, e.g., for lattice simulations,
other powerful nonperturbative techniques are urgently needed.

In this work, we investigate this class of chirally symmetric fermion models
with left/right asymmetry by means of the functional RG. The functional RG and
associated RG flow equations such as the Wetterich equation
\cite{Wetterich:1993yh} are an appropriate tool for this purpose, since
fermions and bosons can both be treated equally well. Also, the flow equation
is an exact equation and facilitates the construction of non-perturbative
approximation schemes, see \cite{Berges:2000ew} for reviews.

After a classification of possible fermionic models in
Sec.~\ref{sec:symmetry}, we concentrate on strong correlations in a scalar
parity-conserving channel. We study the critical behavior of a possible
condensation of bosonic bi-fermion composites into this channel. The resulting
effective scalar-fermion Yukawa model is reminiscent to the standard-model
Higgs-Yukawa sector. We analyze this effective Yukawa model in
Sec.~\ref{sec:RGflow} with the aid of the functional RG, deriving the
nonperturbative RG flow equations to next-to-leading order in a derivative
expansion.  

The left/right asymmetry of our model is controlled by varying the number
$\NL$ of left-handed fermions. This imbalance provides us with an external
control parameter for the relative amplitude of boson and fermion
fluctuations. This allows us to vary the fixed-point properties of the RG
flow, implying also a variation of the critical properties of these
systems. In particular, the fixed-point potential for the bosonic order
parameter can be in the symmetric regime for small $\NL$ or in the broken
regime for larger $\NL$, as discussed in Sec.~\ref{sec:FP}. The small $\NL$
regime turns out to be particularly interesting, as the anomalous dimensions
of all fields have to satisfy a sum rule within our truncation, in order to
give rise to a nontrivial fixed point in the Yukawa coupling. 

The quantitative reliability of our results for the critical properties can be
checked by a reduction of our model to a corresponding purely bosonic O$(N)$
model. For the latter, our method has exhaustively been investigated and used
with a remarkable quantitative success
\cite{Berges:2000ew,Litim:2002cf,Benitez:2009xg}; the results for critical
exponents have reached an accuracy which is comparable if not superior to that
of, e.g., lattice simulations and high-order $\epsilon$ expansions, see
\cite{Benitez:2009xg} for a recent comparison. Our inclusion of fermions can
reliably be based on this footing, as our models approach an O($2\NL$) model
in the limit of large $\NL$; this is because the composite scalar degrees of
freedom dominate the fluctuation contributions as a result of the chiral
structure. For smaller $\NL$, our results show quantitative as well as
qualitative differences to O($N$) models, as expected due to the fermionic
contributions. The resulting models and quantitative findings constitute and
characterize a new set of universality classes, classified by the chiral
symmetry content. In particular, we provide for quantitative predictions for
the critical exponents for these universality classes, which have not been
investigated with any other method so far.  We believe that these can serve as
a first benchmark for other nonperturbative methods which are urgently needed
for a study of chiral phase transitions in strongly correlated chiral
fermions.

\section{Classical action and symmetry transformations}
\label{sec:symmetry}

Let us first consider a general fermionic model in $d=2+1$ Euclidean
dimensions with local quartic self-interaction, being invariant under chiral
$\mathrm U(N_\mathrm L)_\mathrm L\otimes \mathrm U(N_\mathrm R)_\mathrm R$
transformations. The Dirac algebra
\begin{equation}
\{\gamma_\mu,\gamma_\nu\} = 2\delta_{\mu\nu},
\end{equation}
could minimally be realized by an irreducible representation in terms of
$2\times 2$ matrices. As this representation does not permit a chiral
symmetry, the massless theory could not be separated from the massive theory
by an order-disorder transition. We therefore work exclusively with a $4\times
4$ \emph{reducible} representation of the Dirac algebra
\begin{equation}
\gamma_\mu=
\begin{pmatrix}
0 & -\ir\sigma_\mu\\
\ir \sigma_\mu & 0\\
\end{pmatrix}, \quad \mu=1,2,3,
\end{equation}
with $\{\sigma_1,\sigma_2,\sigma_3\}$ being the $2\times2$ Pauli
matrices. Such models with four-component fermions in three dimensions have
extensively been studied in the past (e.g.\ in the context of spontaneous chiral
symmetry breaking in $\textrm{QED}_3$ or the Thirring model)
\cite{Pisarski:1984dj,Appelquist:1985vf,Gomes:1990ed}.
There has been renewed interest in the last years, due to potential
applications to high-$T_c$ superconductivity in cuprates or electronic
properties of graphene
\cite{Maris:1996zg,Hands:2002dv,Herbut:2009qb}. For a review,
see e.g., \cite{Mavromatos:2003ss}. There are now \emph{two}
other $4\times4$ matrices which anticommute with all $\gamma_\mu$ as well as
with each other,
\begin{equation}
\gamma_4=
\begin{pmatrix}
0 & \mathbbm{1}\\
\mathbbm{1} & 0\\
\end{pmatrix} \quad \text{and} \quad
\gamma_5=\gamma_1\gamma_2\gamma_3\gamma_4=
\begin{pmatrix}
\mathbbm{1} & 0\\
0 & -\mathbbm{1}\\
\end{pmatrix}.
\end{equation}
Together with
\begin{equation}
\mathbbm{1}, \ \sigma_{\mu\nu}:=\frac{\ir}{2}[\gamma_\mu,\gamma_\nu] \
(\mu<\nu), \ \ir\gamma_\mu\gamma_4, \ \ir\gamma_\mu\gamma_5, \
\ir\gamma_4\gamma_5,
\end{equation}
these $16$ matrices form a complete basis of the $4\times 4$ Dirac algebra,
\begin{equation}
\left\{\gamma_A\right\}_{A=1,\dots,16} =
\left\{\mathbbm{1},\gamma_\mu,\gamma_4,\sigma_{\mu\nu},\ir\gamma_\mu\gamma_4,
\ir\gamma_\mu\gamma_5,\ir\gamma_4\gamma_5,\gamma_5\right\}.
\end{equation}
Now, we define chiral projectors
\begin{equation}
P_\mathrm{L/R} = \frac{1}{2}(\mathbbm 1\pm \gamma_5)
\end{equation}
that allow us to decompose a Dirac fermion $\psi$ into the left- and right-handed Weyl
spinors $\psi_\mathrm L$ and $\psi_\mathrm R$,
\begin{equation}
\psi_\mathrm{L/R} = P_\mathrm{L/R} \psi, \quad \bar\psi_\mathrm{L/R} = \bar\psi P_\mathrm{R/L}.
\end{equation}
($\psi$ and $\bar\psi$ are considered as independent field variables in our
Euclidean formulation.) Note that there is a certain freedom of choice of the
notion of chirality here: we could have chosen just as well \mbox{$\tilde
  P_\mathrm{L/R} = (\mathbbm 1\pm\gamma_4)/2$} or \mbox{$\hat P_\mathrm{L/R} =
  (\mathbbm 1\pm\ir\gamma_4\gamma_5)/2$} as chiral projectors. This would
have led us to different definitions of the decomposition into Weyl
spinors. All these chiralities remain conserved under Lorentz
transformations since all three projectors commute with the generators of the
Lorentz transformation of the Dirac spinors, $[\gamma_5,\sigma_{\mu\nu}] =
[\gamma_4,\sigma_{\mu\nu}] = [\ir\gamma_4\gamma_5,\sigma_{\mu\nu}] = 0$.

We consider $N_\mathrm R$ right-handed and $N_\mathrm L$ left-handed fermions,
where $N_\mathrm R$ and $N_\mathrm L$ do not have to be identical. We impose a
chiral $\mathrm U(N_\mathrm L)_\mathrm L\otimes \mathrm U(N_\mathrm R)_\mathrm
R$ symmetry with corresponding field transformations which act independently
on left- and right-handed spinors,
\begin{align}
\mathrm U(N_\mathrm L)_\mathrm L:\quad \psi^a_\mathrm L &\mapsto
U^{ab}_\mathrm L \psi^b_\mathrm L, \quad \bar\psi^a_\mathrm L \mapsto
\bar\psi^b_\mathrm L (U_\mathrm L^\dagger)^{ba}, \\ 
\mathrm U(N_\mathrm R)_\mathrm R:\quad \psi^a_\mathrm R &\mapsto
U^{ab}_\mathrm R \psi^b_\mathrm R, \quad \bar\psi^a_\mathrm R \mapsto
\bar\psi^b_\mathrm R (U_\mathrm R^\dagger)^{ba}. 
\end{align}
Here, $U_\mathrm L$ and $U_\mathrm R$ are unitary $N_\mathrm L\times
N_\mathrm L$ and $N_\mathrm R\times N_\mathrm R$ matrices, respectively. For
$U^{ab}_\mathrm L = \er^{\ir\alpha}\delta^{ab}$ and $U^{ab}_\mathrm R =
\er^{-\ir\alpha}\delta^{ab}$ we obtain the usual $\mathrm U(1)_\mathrm A$ axial
transformations, whereas for $U_\mathrm L^{ab}=\er^{\ir\alpha}\delta^{ab}$ and
$U_\mathrm R^{ab}=\er^{\ir\alpha}\delta^{ab}$ we get $\mathrm U(1)_\mathrm
V$ phase rotations. The symmetry thus is
\begin{equation}
\mathrm U(N_\mathrm L)_\mathrm L\otimes \mathrm U(N_\mathrm R)_\mathrm R\cong
\mathrm {SU}(N_\mathrm L)_\mathrm L \otimes \mathrm {SU}(N_\mathrm R)_\mathrm R
\otimes \mathrm U(1)_\mathrm A \otimes \mathrm U(1)_\mathrm V,
\end{equation}
with chiral SU($N_{\text{L,R}}$) factors. 

Due to the reducible representation of the Dirac algebra, there is also some
freedom in the definition of the discrete transformations \cite{Gomes:1990ed}.
Charge conjugation may be
implemented by either
\begin{equation}
\mathcal C: \, \psi^a_\mathrm{L/R} \mapsto \left(\bar\psi^a_\mathrm{L/R}
C\right)^T, \, \bar\psi^a_\mathrm{L/R} \mapsto -\left(C^\dagger
\psi^a_\mathrm{L/R}\right)^T,
\end{equation}
with $C = \gamma_2\gamma_5$, or
\begin{equation}
\mathcal{\tilde C}: \, \psi^a_\mathrm{L/R} \mapsto
\left(\bar\psi^a_\mathrm{R/L}
\tilde C\right)^T, \, \bar\psi^a_\mathrm{L/R} \mapsto -\left(\tilde C^\dagger
\psi^a_\mathrm{R/L}\right)^T,
\end{equation}
with $\tilde C =\gamma_2\gamma_4$, or a unitary combination thereof. In the same manner, the parity transformation
corresponding to
\begin{equation}
(x_1,x_2,x_3) \mapsto (-x_1,x_2,x_3) =: \tilde x,
\end{equation}
with $(x_1,x_2)$ as space coordinates and $x_3$ as the (Euclidean) time
coordinate, may be implemented by either
\begin{equation}
\mathcal P: \, \psi^a_\mathrm{L/R}(x) \mapsto P
\psi^a_\mathrm{L/R}(\tilde x), \, \bar\psi^a_\mathrm{L/R}(x) \mapsto
\bar\psi^a_\mathrm{L/R}(\tilde x) P^\dagger, \,
\end{equation}
with $P = \gamma_1\gamma_4$, or
\begin{equation}
\mathcal{\tilde P}: \, \psi^a_\mathrm{L/R}(x) \mapsto \tilde P
\psi^a_\mathrm{R/L}(\tilde x), \, \bar\psi^a_\mathrm{L/R}(x) \mapsto
\bar\psi^a_\mathrm{R/L}(\tilde x) {\tilde P}^\dagger,
\end{equation}
with $\tilde P = \gamma_1\gamma_5$. Similarly, time reversal corresponding to
\begin{equation}
(x_1,x_2,x_3) \mapsto (x_1,x_2,-x_3) =: \hat x
\end{equation}
reads either
\begin{equation}
\mathcal{T}: \, \psi^a_\mathrm{L/R}(x) \mapsto T
\psi^a_\mathrm{L/R}(\hat x),\,
\bar\psi^a_\mathrm{L/R}(x)\mapsto\bar\psi^a_\mathrm{L/R}(\hat x) {T}^\dagger,
\end{equation}
with $T=\gamma_2\gamma_3$, or
\begin{equation}
\mathcal{\tilde T}: \, \psi^a_\mathrm{L/R}(x) \mapsto \tilde T
\psi^a_\mathrm{R/L}(\hat x),\,
\bar\psi^a_\mathrm{L/R}(x)\mapsto\bar\psi^a_\mathrm{R/L}(\hat x) {\tilde
T}^\dagger,
\end{equation}
with $\tilde T=\gamma_1$. Note that every matrix $\in\{C,\tilde C,P,\tilde P,T,\tilde T\}$
is unitary, such that charge conjugation and parity inversion are unitary,
and time reversal is anti-unitary.

In order to derive the explicit transformation properties of the bilinears, it
is useful to recall that $\gamma_1$ and $\gamma_3$ are antisymmetric and purely
imaginary, whereas $\gamma_2$, $\gamma_4$, and $\gamma_5$ are symmetric and
real. The results are listed in Table \ref{tab:CPT_Transformation_1}, where we
have introduced
\begin{eqnarray}
\tilde\gamma_\mu &:=& (-\gamma_1,\gamma_2,\gamma_3)_\mu, \\ 
(\tilde \sigma_{12},\tilde \sigma_{13},\tilde\sigma_{23})&:=&(-\sigma_{12},-\sigma_{13},\sigma_{23}),\\
\hat\gamma_\mu &:=& (\gamma_1,\gamma_2,-\gamma_3)_\mu, \\
(\hat \sigma_{12},\hat \sigma_{13},\hat\sigma_{23})&:=&(\sigma_{12},-\sigma_{13},-\sigma_{23}).
\end{eqnarray}
\begin{table*}[tbp]
\caption{\label{tab:CPT_Transformation_1}Properties of fermion bilinears under
  discrete transformations. The arguments of the transformed fields are
  $\tilde x = (-x_1,x_2,x_3)$ in the case of parity and $\hat
  x=(x_1,x_2,-x_3)$ in the case of time reversal. The bilinears with $(\mathrm
  L \leftrightarrow \mathrm R)$ transform analogously.}
\begin{ruledtabular}\begin{tabular}{>{$}l<{$}*{6}{>{$}c<{$}}}
& \mathcal C & \mathcal{\tilde C} & \mathcal P & \mathcal{\tilde P} & \mathcal T
& \mathcal{\tilde T} \tabularnewline \hline
\bpsiL^a\psiR^b & \bpsiR^b\psiL^a & \bpsiL^b\psiR^a & \bpsiL^a\psiR^b &
\bpsiR^a\psiL^b & \bpsiL^a\psiR^b & \bpsiR^a\psiL^b
\tabularnewline
\bpsiL^a\gamma_\mu\psiL^b & -\bpsiL^b\gamma_\mu\psiL^a &
-\bpsiR^b\gamma_\mu\psiR^a & \bpsiL^a\tilde\gamma_\mu\psiL^b &
\bpsiR^a\tilde\gamma_\mu\psiR^b & -\bpsiL^a\hat\gamma_\mu\psiL^b &
-\bpsiR^a\hat\gamma_\mu\psiR^b
\tabularnewline
\bpsiL^a\sigma_{\mu\nu}\psiR^b & -\bpsiR^b\sigma_{\mu\nu}\psiL^a &
-\bpsiL^b\sigma_{\mu\nu}\psiR^a & \bpsiL^a\tilde\sigma_{\mu\nu}\psiR^b &
\bpsiR^a\tilde\sigma_{\mu\nu}\psiL^b & -\bpsiL^a\hat\sigma_{\mu\nu}\psiR^b &
-\bpsiR^a\hat\sigma_{\mu\nu}\psiL^b
\tabularnewline
\bpsiL^a\gamma_4\psiL^b & \bpsiL^b\gamma_4\psiL^a & -\bpsiR^b\gamma_4\psiR^a &
-\bpsiL^a\gamma_4\psiL^b & \bpsiR^a\gamma_4\psiR^b & \bpsiL^a\gamma_4\psiL^b &
-\bpsiR^a\gamma_4\psiR^b
\tabularnewline
\bpsiL^a\ir\gamma_\mu\gamma_4\psiR^b & \bpsiR^b\ir\gamma_\mu\gamma_4\psiL^a &
-\bpsiL^b\ir\gamma_\mu\gamma_4\psiR^a &
-\bpsiL^a\ir\tilde\gamma_\mu\gamma_4\psiR^b &
\bpsiR^a\ir\tilde\gamma_\mu\gamma_4\psiL^b &
\bpsiL^a\ir\hat\gamma_\mu\gamma_4\psiR^b &
-\bpsiR^a\ir\hat\gamma_\mu\gamma_4\psiL^b
\end{tabular}\end{ruledtabular}
\end{table*}
These transformation properties facilitate a discussion of possible bilinears
and 4-fermi terms in the action of our model. In addition to Lorentz
invariance, we impose an invariance of our theory under $\mathrm
U(N_\mathrm L)_\mathrm L\otimes \mathrm U(N_\mathrm R)_\mathrm R$ chiral
transformations, $\mathcal{C}$ charge conjugation, $\mathcal{P}$ parity
inversion, and $\mathcal{T}$ time reversal.  The theory then automatically is
also invariant under $\mathcal{\tilde C \tilde P}$, $\mathcal{\tilde P \tilde
  T}$, and $\mathcal{\tilde C\tilde T}$ transformations, since
$\mathcal{\tilde C\tilde P}=\mathcal{CP}$, $\mathcal{\tilde P \tilde
  T}=\mathcal{PT}$, and $\mathcal{\tilde C \tilde T}=\mathcal{CT}$.
(The equivalence holds up to $\mathrm U(1)_\mathrm V$ phase rotations of the
spinors.) As a consequence, no bilinears to zeroth order in derivatives are
permitted. To first order, only the standard chiral kinetic terms
\begin{equation}
\bpsiL^a\ir\partial_\mu\gamma_\mu\psiL^a \quad \text{and} \quad
\bpsiR^a \ir\partial_\mu\gamma_\mu\psiR^a
\end{equation}
can appear. In particular, all possible mass terms are excluded by symmetry:
$\bpsiL^a\psiR^a$ and $\bpsiR^a\psiL^a$ are not chirally symmetric, and
$\bpsiL^a\gamma_4\psiL^a$ as well as $\bpsiR^a\gamma_4\psiR^a$ are not
invariant under $\mathcal{P}$ parity transformations. The same holds for terms
involving $\ir\gamma_4\gamma_5$. On the level of 4-fermi operators, the
interaction terms must have the form
\begin{gather}\label{eq:all4fermi_A}
\left(\bpsiL^a\gamma_A\psiL^b\right)\left(\bpsiL^b\gamma_A\psiL^a\right),
\quad
\left(\bpsiR^a\gamma_A\psiR^b\right)\left(\bpsiR^b\gamma_A\psiR^a\right)
\end{gather}
with $\gamma_A\in\{\gamma_\mu,\gamma_4\}$,
\begin{gather}
\left(\bpsiL^a\gamma_B\psiR^b\right)\left(\bpsiR^b\gamma_B\psiL^a\right)
\quad\text{with}\quad
\gamma_B\in\{\mathbbm{1},\ir\gamma_\mu\gamma_4\},\label{eq:all4fermi_B} 
\end{gather}
or, with inverse flavor structure,
\begin{gather}\label{eq:4fermi_inverseFlavor}
\left(\bpsiL^a\gamma_A\psiL^a\right)\left(\bpsiL^b\gamma_A\psiL^b\right)
,\quad
\left(\bpsiR^a\gamma_A\psiR^a\right)\left(\bpsiR^b\gamma_A\psiR^b\right),\nonumber\\
\left(\bpsiR^a\gamma_A\psiR^a\right)\left(\bpsiL^b\gamma_A\psiL^b\right).
\end{gather}
Terms with $\gamma_A\in\{\ir\gamma_\mu\gamma_5,\ir\gamma_4\gamma_5\}$ or
$\gamma_B\in\{\gamma_5,\sigma_{\mu\nu}\}$ are equal to these up to a possible
sign, since $\psiL$ and $\psiR$ are eigenvectors of $\gamma_5$, and
$\sigma_{\mu\nu}=-\ir\epsilon_{\mu\nu\rho}\gamma_\rho\gamma_4\gamma_5$. Terms
with $\gamma_A\in\{\mathbbm
1,\gamma_5,\ir\gamma_\mu\gamma_4,\sigma_{\mu\nu}\}$ or
$\gamma_B\in\{\gamma_\mu,\ir\gamma_\mu\gamma_5,\gamma_4,
\ir\gamma_4\gamma_5\}$ are identically zero, since $P_\mathrm R P_\mathrm
L=P_\mathrm L P_\mathrm R=0$. The terms in \Eqref{eq:4fermi_inverseFlavor} are
not independent of the terms in Eqs.~\eqref{eq:all4fermi_A},
\eqref{eq:all4fermi_B}, but are related by Fierz transformations:
\begin{widetext}
\begin{align}
\left(\bpsiL^a\gamma_\mu\psiL^a\right)\left(\bpsiL^b\gamma_\mu\psiL^b\right) & = \frac{1}{2}\left(\bpsiL^a\gamma_\mu\psiL^b\right)\left(\bpsiL^b\gamma_\mu\psiL^a\right) + \frac{3}{2}\left(\bpsiL^a\gamma_4\psiL^b\right)\left(\bpsiL^b\gamma_4\psiL^a\right), \\
\left(\bpsiL^a\gamma_4\psiL^a\right)\left(\bpsiL^b\gamma_4\psiL^b\right) & = \frac{1}{2}\left(\bpsiL^a\gamma_\mu\psiL^b\right)\left(\bpsiL^b\gamma_\mu\psiL^a\right) - \frac{1}{2}\left(\bpsiL^a\gamma_4\psiL^b\right)\left(\bpsiL^b\gamma_4\psiL^a\right),
\end{align}
\begin{align}
\left(\bpsiR^a\gamma_\mu\psiR^a\right)\left(\bpsiL^b\gamma_\mu\psiL^b\right) & = -\frac{3}{2}\left(\bpsiR^a\psiL^b\right)\left(\bpsiL^b\psiR^a\right) - \frac{1}{2}\left(\bpsiR^a\ir\gamma_\mu\gamma_4\psiL^b\right)\left(\bpsiL^b\ir\gamma_\mu\gamma_4\psiR^a\right), \\
\left(\bpsiR^a\gamma_4\psiR^a\right)\left(\bpsiL^b\gamma_4\psiL^b\right) & =
-\frac{1}{2}\left(\bpsiR^a\psiL^b\right)\left(\bpsiL^b\psiR^a\right) +
\frac{1}{2}\left(\bpsiR^a\ir\gamma_\mu\gamma_4\psiL^b\right)\left(\bpsiL^b\ir\gamma_\mu\gamma_4\psiR^a\right). 
\end{align}
\end{widetext}
Here, we have suppressed the analogous equations with $(\mathrm L
\leftrightarrow \mathrm R)$ for simplicity.  We thus end up with six
independent 4-fermi terms preserving $\mathrm U(N_\mathrm L)_\mathrm L\otimes
\mathrm U(N_\mathrm R)_\mathrm R$ chiral and $\mathcal C$, $\mathcal P$, and
$\mathcal T$ symmetry,
\begin{gather}
\left(\bpsiL^a\psiR^b\right)\left(\bpsiR^b\psiL^a\right),  \label{eq:scalar}\\
\left(\bpsiL^a\gamma_4\psiL^b\right)\left(\bpsiL^b\gamma_4\psiL^a\right),  \quad
\left(\bpsiR^a\gamma_4\psiR^b\right)\left(\bpsiR^b\gamma_4\psiR^a\right), \label{eq:pseudoscalar}\\
\left(\bpsiL^a\gamma_\mu\psiL^b\right)\left(\bpsiL^b\gamma_\mu\psiL^a\right),  \quad
\left(\bpsiR^a\gamma_\mu\psiR^b\right)\left(\bpsiR^b\gamma_\mu\psiR^a\right), \label{eq:vector}\\
\left(\bpsiL^a\ir\gamma_\mu\gamma_4\psiR^b\right)\left(\bpsiR^b\ir\gamma_\mu\gamma_4\psiL^a\right). \label{eq:pseudovector}
\end{gather}
Note that the corresponding set in $d=3+1$ dimensions would be smaller as the
$\gamma_4$ terms would be constrained by a larger Lorentz symmetry.

In a (partially) bosonized language after a Hubbard-Stratonovich
transformation, we encounter six boson-fermion interactions: The first one
corresponding to \Eqref{eq:scalar} couples the fermions to a scalar boson
(scalar with respect to $\mathcal{P}$~parity), the second and the third
\eqref{eq:pseudoscalar} to a pseudo-scalar boson, the fourth and the fifth
\eqref{eq:vector} to a vector boson, and the sixth \eqref{eq:pseudovector}
to a pseudo-vector boson. Further bosonic structures in the flavor-singlet
channels appear in the corresponding Fierz transforms of
Eqs.~\eqref{eq:scalar}-\eqref{eq:pseudovector}. 

We expect that a general model based on these interactions exhibits a rich
phase structure, being controlled by the relative strength of the various
interaction channels. Aiming at an analogue of the electroweak phase
transition, we focus in this work on the $\mathcal{P}$~parity-conserving and
Lorentz-invariant condensation channel, parameterized in terms of the first
interaction term. This channel is also invariant under $\mathcal{T}$ time
reversal, whereas the boson
transforms into its complex conjugate under $\mathcal C$ charge
conjugation. Moreover, we confine ourselves to the case of $N_\mathrm R=1$
right-handed fermion flavor and $N_\mathrm L\geq 1$ left-handed fermion
flavors, allowing for a left/right asymmetry similar to the standard model of
particle physics (where $\NL=2$). The microscopic action of our model in the
purely fermionic language then
reads
\begin{eqnarray}
S_\text{4-fermi} &=& \int \mathrm d^3x\big\{\bpsiL^a\ir \dslash \psiL^a +
\bpsiR\ir \dslash \psiR \nonumber\\
&&\quad\quad + 2\lambda\left(\bpsiL^a\psiR\right)\left(\bpsiR\psiL^a\right)\big\}.
\end{eqnarray}
Via Hubbard-Stratonovich transformation, we obtain the equivalent Yukawa action
\begin{eqnarray}
S_\text{Yuk} &=& \int \mathrm d^3x\big\{\frac{1}{2\lambda}
\phi^{a\dagger}\phi^a + \bar\psi^a_\mathrm L\ir \dslash \psi^a_\mathrm L +
\bar\psi_\mathrm R\ir \dslash \psi_\mathrm R \nonumber\\
&&\quad\quad +
\phi^{a\dagger}\bar{\psi}_\mathrm R\psi_\mathrm L^a -
\phi^{a}\bar{\psi}_\mathrm L^a\psi_\mathrm R\big\}, 
\label{eq:Yuk}
\end{eqnarray}
where the complex scalar $\phi^a$ serves as an auxiliary field. The purely
fermionic model can be recovered by use of the algebraic equations of motion for
$\phi^a$ and $\phi^{a\dagger}$,
\begin{align}
\phi^{a} = - 2\lambda\bar\psi_\mathrm R\psi^a_\mathrm L, \quad \phi^{a\dagger}
= 2\lambda\bar\psi^a_\mathrm L\psi_\mathrm R.
\end{align}
Here, we can read off the transformation properties of the scalar field under
the chiral symmetry,
\begin{equation}
\phi^a \mapsto U^{ab}_\mathrm L \phi^b U_\mathrm R ^\dagger, \quad
\phi^a{}^\dagger \mapsto U_\mathrm R \phi^b{}^\dagger (U_\mathrm L ^\dagger)^{ba}.
\end{equation}
The composite scalar field $\phi^a$ represents an order parameter for an
order-disorder transition. As long as $\phi^a$ has a vanishing expectation
value, the system is in the symmetric phase with full chiral
U($\NL$)$_\mathrm{L} \times \mathrm U(1)_{\mathrm{R}}$ symmetry; the fermions
are
massless, whereas the scalars are generically massive as determined by the
symmetry-preserving effective potential for the scalars. If $\phi^a$ acquires
a vacuum expectation value the chiral SU($\NL$) factor is broken down to a
residual SU($\NL-1$) symmetry. In addition, the axial U(1)$_\text{A}$ is
broken, whereas the charge-conserving vector U(1)$_\text{V}$ is
preserved. In the broken phase, the spectrum
consists of one massive Dirac fermion, one massive bosonic radial mode,
$\NL-1$ massless left-handed Weyl fermions and $2\NL -1$ massless Goldstone
bosons.

Near the phase transition, we expect the order-parameter fluctuations to
dominate the critical behavior of the system. Universality suggests that the
degrees of freedom parameterized by the action \eqref{eq:Yuk} are sufficient to
quantify the critical behavior of this transition, independently of the
presence of further microscopic fermionic interactions of
Eqs.~\eqref{eq:scalar}-\eqref{eq:pseudovector}. Concentrating on the action
\eqref{eq:Yuk}, we observe that the purely scalar sector, i.e., the scalar
mass term, has a larger symmetry group of O$(2\NL)$-type. It is therefore
instructive to compare the critical behavior of our fermionic model with that
of a standard scalar O$(2\NL)$ model which is known to undergo a second order
phase transition associated with a Wilson-Fisher fixed point. Differences in
the corresponding critical behaviors can then fully be attributed to fermionic
fluctuations near the phase transition.

\section{Effective average action and RG flow}
\label{sec:RGflow}

Integrating out fluctuations momentum shell by momentum shell near the phase
transition described above, the effective action (effective Wilson-type
Hamiltonian) is expected to acquire all possible operators of mixed scalar and
fermionic nature compatible with the symmetries of the action
\eqref{eq:Yuk}. In the present work, we constrain the flow of this action
functional to lie in the subspace of the full theory space spanned by the
following ansatz valid at an RG scale $k$:
\begin{eqnarray}\label{eq:SULtruncation}
\Gamma_k&=&\int \mathrm d^dx\Big\{
Z_{\mathrm{L},k}\bar{\psi}_{\mathrm{L}}^a\ir\dslash\psi_{\mathrm{L}}^a
+Z_{\mathrm{R},k}\bar{\psi}_{\mathrm{R}}\ir\dslash\psi_{\mathrm{R}}\nonumber\\
&{}&+Z_{\phi, k}\left(\partial_{\mu}\phi^{a\dagger}\right)\left(\partial^{\mu}\phi^a\right)
+U_k(\phi^{a\dagger}\phi^a)\nonumber\\
&{}&+\bar h_k\bar{\psi}_{\mathrm{R}}\phi^{a\dagger}\psi_{\mathrm{L}}^a-\bar
h_k\bar{\psi}_{\mathrm{L}}^a\phi^{a}\psi_{\mathrm{R}}\Big\}. 
\end{eqnarray}
The fermion fields $\psi_{\mathrm{L}}^a$ and $\psi_{\mathrm{R}}$ have standard
kinetic terms but can pick up different wave function renormalizations
$Z_{\mathrm{L},k}$ and $Z_{\mathrm{R},k}$. The index $a$ runs from $1$ to $\NL
$. The bosonic sector involves a standard kinetic term with wave function
renormalization $Z_{\phi , k}$ and an effective potential $U_k(\rho)$, where
$\rho=\phi^{a\dagger}\phi^a$. It is sometimes useful, to express the complex
scalar field in terms of a real field basis by defining
\begin{eqnarray}
 \phi^a = \frac{1}{\sqrt{2}}(\phi_1^a + \ir \phi_2^a),\quad \phi^{a\dagger} =
 \frac{1}{\sqrt{2}}(\phi_1^a - \ir \phi_2^a)\,, 
\end{eqnarray}
where $\phi_1^a, \phi_2^a\in\mathbbm{R}$. All parameters in the effective
average action are understood to be scale dependent, which is indicated by the
momentum-scale index $k$. The scale dependence is governed by the Wetterich
equation, which allows for a nonperturbative construction of quantum field theory in
terms of the effective average action $\Gamma_k$ \cite{Wetterich:1993yh}:
\begin{equation}\label{flowequation}
\partial_t\Gamma_k[\Phi]
=\frac{1}{2}\mathrm{STr}\left\{\left[\Gamma^{(2)}_k[\Phi]+R_k\right]^{-1}
\left(\partial_tR_k\right)\right\}.
\end{equation}
Here, $\Gamma^{(2)}_k[\Phi]$ is the second functional derivative with respect
to the field $\Phi$, the latter representing a collective field variable for
all bosonic or fermionic degrees of freedom, and $R_k$ denotes a
momentum-dependent regulator function that suppresses IR modes below a
momentum scale $k$. The solution to the Wetterich equation provides an RG
trajectory in theory space, interpolating between the bare action
$S_\mathrm{Yuk}$ to be quantized $\Gamma_{k\to\Lambda}\to S_\mathrm{Yuk}$ and
the full quantum effective action $\Gamma=\Gamma_{k\to 0}$, being the
generating functional of 1PI correlation functions. For reviews, see
e.g., \cite{Berges:2000ew}.

The ansatz \eqref{eq:SULtruncation}, in fact, represents the next-to-leading
order in a systematic derivative expansion of the effective potential in the
scalar sector and a leading-order vertex expansion in the fermionic
sector. (The leading-order derivative expansion is obtained by setting all
wave function renormalizations $Z_{(\phi,\mathrm{L,R}),k}=\mathrm{const.}$)
This expansion can consistently be extended to higher orders and thus defines
a legitimate and controllable nonperturbative approximation scheme. It has
indeed proved its quantitative reliability already in a number of examples
involving Yukawa sectors
\cite{Rosa:2000ju,Jungnickel:1995fp,Braun:2009si,Birse:2004ha}.

In order to fix the standard RG invariance of field rescalings, we define the
renormalized fields as
\begin{equation}
\tilde{\phi}=Z_{\phi,k}^{1/2}\phi,\quad
\tilde{\psi}_{\mathrm{L/R}}=Z_{\mathrm{L/R},k} ^{1/2}\psi_{\mathrm{L,R}}.
\end{equation}
For the search for a fixed point where the system is scale invariant, it is
useful to introduce dimensionless renormalized quantities. In order to display
the dimension dependence, we perform the analysis in $d$ spacetime dimensions,
where the dimensionless renormalized field, Yukawa coupling and scalar
potential read
\begin{eqnarray}\label{eq:dimensionless}
\tilde{\rho}&=&Z_{\phi,k}k^{2-d}\rho,\\
 h_k^2&=&Z_{\phi,k}^{-1}Z_{\mathrm{L},k}^{-1}Z_{\mathrm{R},k}^{-1}k^{d-4}\bar h_k^2 ,\\
 u_k(\tilde\rho)&=&k^{-d}U_k(\rho)|_{\rho=k^{d-2}\tilde\rho/Z_{\phi,k}}.
\end{eqnarray}
The flow of the wave function renormalizations $Z_{\phi , k},
Z_{\mathrm{L},k}$ and $Z_{\mathrm{R},k}$ can be expressed in terms of
scale-dependent anomalous dimensions
\begin{equation}
	\eta_{\phi}=-\partial_t \mbox{ln} Z_{\phi,k},\quad
        \eta_{\mathrm{L/R}}=-\partial_t \mbox{ln} Z_{\mathrm{L/R},k}\,.
\end{equation}
With these preliminaries, we can compute the flow
of the effective potential (see App. \ref{sec:flowequations}),
\begin{eqnarray}\label{eq:potflow}
\partial_tu_k&=&-du_k+\tilde{\rho}(d-2+\eta_{\phi})u_k'+2v_d\Bigl[
(2N_{\mathrm{L}}-1)l_0^d(u_k')\nonumber\\
&&+l_0^d(u_k'+2\tilde{\rho}
u_k'')-d_{\gamma}(N_{\mathrm{L}}-1)l_{0,\mathrm L}^{(\mathrm
F)d}(0)\nonumber\\
&&
-d_{\gamma}l_{0,\mathrm L}^{(\mathrm F)d}
(h_k^2\tilde{\rho})-d_{\gamma}l_{0,\mathrm R}^{(\mathrm
F)d}(h_k^2\tilde{\rho}) \Bigr],
\end{eqnarray}
where $v_d^{-1}=2^{d+1}\pi^{d/2}\Gamma(d/2)$, $d_{\gamma}=4$ is
the dimension of the $\gamma$ matrices, and the threshold functions,
\begin{eqnarray}
l^d_n(\omega) &=& \frac{2(\delta_{n,0}+n)}{d}\Big(1
-\frac{\eta_\phi}{d+2}\Big) \frac{1}{(1+\omega)^{n+1}}, \label{eq:threshold1}\nonumber\\
l^{(\mathrm{F})d}_{n,\mathrm{L/R}}(\omega) &=& \frac{2(\delta_{n,0}+n)}{d}\Big(1
-\frac{\eta_{\mathrm{L/R}}}{d+1}\Big) \frac{1}{(1+\omega)^{n+1}},
\end{eqnarray}
encode the information about a possible decoupling of massive modes. For the
momentum regularization, we have used a linear regulator function $R_k$ here
which is optimized for the derivative truncation \cite{Litim:2001up}.

Whereas the flow of the Yukawa coupling is unambiguous in the symmetric
regime, the Goldstone and radial modes can generally develop different
couplings in the broken regime. Here, we concentrate on the Goldstone-mode
Yukawa coupling to the fermions, as the radial mode becomes massive and
decouples in the broken regime. In both phases, the flow of the Yukawa
coupling can be written as (see App. \ref{sec:flowequations})
\begin{eqnarray}\label{eq:hquadratFlow}
\partial_th_k^2&=&(\eta_{\phi}+\eta_\mathrm L+\eta_\mathrm
R+d-4)h_k^2\\&&-8v_dh_k^4\kappa_ku_{k} ''l_{111}^{(\mathrm{FB})d}(\kappa_kh_k^2,
u_{k}'+2\kappa_ku_{k}'', u_{k}'),\nonumber
\end{eqnarray}
where the scalar-potential terms have to be evaluated on the $k$-dependent
minimum of the potential $\tilde\rho_{\text{min}}\equiv \kappa_k$. In the
symmetric regime, we have, of course, $\kappa_k=0$. The threshold function
occurring in \Eqref{eq:hquadratFlow} reads for the linear regulator:
\begin{widetext}
\begin{multline*}
l_{n_1,n_2,n_3}^{(\mathrm{FB})d}(\omega_1,\omega_2,\omega_3)=\frac{2}{d}\frac{1}{(1+\omega_1)^{n_1}(1+\omega_2)^{n_2}(1+\omega_3)^{n_3}}\times \\
\left[ \frac{n_1}{1+\omega_1}\left(
1-\frac{\frac{1}{2}(\eta_{\mathrm{L}}+\eta_{\mathrm{R}})}{d+1} \right)
+\frac{n_2}{1+\omega_2}\left( 1-\frac{\eta_{\phi}}{d+2} \right)
+\frac{n_3}{1+\omega_3}\left( 1-\frac{\eta_{\phi}}{d+2} \right) \right].
\end{multline*}
Setting the anomalous dimensions to zero defines the leading-order derivative
expansion. At next-to-leading order, it is important to distinguish
between $Z_{\mathrm{L},k}$ and $Z_{\mathrm{R},k}$ as they acquire different loop
contributions, see below. The flows of the anomalous dimensions read (see App.
\ref{sec:flowequations})
\begin{align}\label{eq:anomalousDimensions}\begin{split}
\eta_{\phi}&=\frac{16v_d}{d}u_{k}''^2\kappa_km_{22}^d(u_{k}'+2\kappa_ku_{k}'',u_
{k}')+\frac{8v_dd_{\gamma}}{d}\left[ \kappa_kh_k^4m_2^{(\mathrm
F)d}(\kappa_kh_k^2)+h_k^2m_4^{(\mathrm F)d}(\kappa_kh_k^2) \right],\\
\eta_\mathrm L&=\frac{8v_d}{d}h_k^2\big[ m_{12}^{(\mathrm{FB})d}(h_k^2\kappa_k,
u_{k}'+2\kappa_ku_{k}'')+m_{12}^{(\mathrm{FB})d}(h_k^2\kappa_k,
u_{k}')\big],\\
\eta_\mathrm R&=\frac{8v_d}{d}h_k^2\left[ m_{12}^{(\mathrm{FB})d}(h_k^2\kappa_k,
u_{k}'+2\kappa_ku_{k}'')+m_{12}^{(\mathrm{FB})d}(h_k^2\kappa_k,
u_{k}') +2(N_{\mathrm{L}}-1)m_{12}^{(\mathrm{FB})d}(0, u_{k}')\right],
\end{split}\end{align}
\end{widetext}
where the potential is again evaluated at $\tilde{\rho}=\kappa_k$. Here, we
have also introduced the corresponding threshold functions for the linear
regulator
\begin{align}\label{eq:etaThresholds}\begin{split}
m^d_{n_1,n_2}(\omega_1,\omega_2)  &=
\frac{1}{(1+\omega)^{n_1}(1+\omega)^{n_2}},\\
m^{(\mathrm{F})d}_{2}(\omega) & = \frac{1}{(1+\omega)^4},\\
m^{(\mathrm{F})d}_{4}(\omega) & =
\frac{1}{(1+\omega)^4}+\frac{1-\eta_\psi}{d-2}\frac{1}{(1+\omega)^3}\\
&-\left(\frac{1-\eta_\psi}{2d-4}+\frac{1}{4}\right)\frac{1}{(1+\omega)^2}, \\
m^{(\mathrm{FB})d}_{n_1,n_2}(\omega_1,\omega_2) & =
\left(1-\frac{\eta_\phi}{d+1}\right)\frac{1}{(1+\omega_1)^{n_1}(1+\omega_2)^{
n_2 }},
\end{split}\end{align}
and $\eta_\psi:=\frac{1}{2}(\eta_{\mathrm{R}}+\eta_{\mathrm{L}})$.

\section{Fixed points and critical exponents}
\label{sec:FP}

The Wetterich equation provides us with the flow of the generalized couplings
$\partial_t g_i= \beta_{i}(g_1,g_2,\dots)$ of our truncation, where the $g_i$
correspond to the Yukawa coupling and, e.g., the expansion coefficients of the
scalar potential, etc. A fixed point $g^\ast$ is defined by
\begin{equation}
\forall i:\quad \beta_i(g_1^{\ast},g_2^{\ast},...)=0.
\end{equation}
If the fixed point separates the disordered from an ordered phase, it is a
candidate for a second-order phase transition.  For the investigation of the
critical properties of the theory near this transition, we consider the
fixed-point regime, where the flow can be linearized around the fixed point,
\begin{equation}
 \partial_t g_i = B_i{}^j (g^\ast_j-g_j)+\dots, \quad B_i{}^j =\frac{\partial
   \beta_{i}}{\partial g_j} \Big|_{g=g^\ast}.\label{eq:lin}
\end{equation}
Let us denote the eigenvalues of the stability matrix $B_i^j$ by
$\omega_I$. The index $I$ labels the order of the eigenvalues according to
their real part, starting with the smallest one which we call
$\omega_0$. Negative $\omega_I$ correspond to RG relevant directions for which
an IR (UV) fixed point is repulsive (attractive). In analogy to the notion of
critical phenomena in $O(N )$-models, we define $\nu = - 1/\omega_0$,
characterizing the critical exponent of the correlation length near the
critical temperature. This is indeed justified, as the largest critical
exponent is associated with the strongest RG relevant direction, being in turn
related to the distance from the critical temperature. Thus, $\nu$, in fact,
corresponds to the standard correlation-length exponent. The subleading
exponent is traditionally called $\omega = \omega_1$.

\begin{figure}[ht]
\centering
\includegraphics[width=0.40\textwidth]{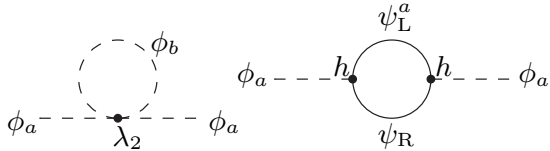}
\caption{Loop contributions
to the
renormalization flow of the mass or the vev.
The left loop involves only inner boson lines. The vertex $\lambda_2$ allows
for a coupling between all available boson components. This implies a linear
dependence on $\NL $ for the renormalization of the boson contribution,
see below. On the right panel, we depict the corresponding fermion loop
contribution. The incoming boson $\phi_a$ fully determines the structure of the
fermion loop and does not allow for other left-handed inner fermions than
$\psi_{\mathrm{L}}^a$, inhibiting an $\NL $ dependence of the algebraic
weight of this loop.}
\label{fig:loops}
\end{figure}

In order to analyze the fixed point structure of this model, we will distinguish
two different regimes of the system namely the symmetric regime (SYM), where the
vacuum expectation value (vev) of the boson field is zero, and the regime of
spontaneously broken symmetry
(SSB), where it is nonzero. As boson and fermion fluctuations generically
contribute with opposite sign, the existence of a fixed point requires a
balancing between both contributions together with potential dimensional
scaling terms (such as, e.g., the first term on the right-hand side of
\Eqref{eq:hquadratFlow}). We observe that this balancing is indeed possible in
both regimes, depending on the number of left-handed fermion flavors. The
origin of this flavor-number dependence is illustrated for the running of the
scalar mass term or vacuum expectation value in Fig.~\ref{fig:loops}.
Whereas the scalar loop carries a weight $\sim\NL$ as all scalar degrees of
freedom can contribute, the left/right asymmetry structure leads to a weight
$\sim \mathcal O(1)$ for the fermion loops in Fig.~\ref{fig:loops}. In the
following, we analyze the different regimes in detail. 

\subsection{The symmetric regime}

Here and in the following, we drop the subscript $k$ indicating the scale
dependence of the running couplings for simplicity. The presence of this scale
dependence will be implicitly understood. Also we restrict our investigations to $d=3$ from now on.

In the symmetric regime, we employ the following expansion of the
effective potential $u$
\begin{eqnarray}\label{eq:symeffpot}
  u\equiv u_k&=&\sum_{n=1}^{N_p}\frac{\lambda_{n}}{n!}\tilde{\rho}^n 
  = m^2\tilde{\rho}+\frac{\lambda_{2}}{2!}\tilde{\rho}^2
  +\frac{\lambda_{3}}{3!}\tilde{\rho}^3+...
\end{eqnarray}
($m^2\equiv \lambda_1$). The minimum is assumed to be at $\tilde\rho=0$,
implying $\kappa=0$. This also implies that $m^2\geq 0$.  This expansion is
plugged into the flow equations for the effective potential, the Yukawa
coupling and the anomalous dimensions. An expansion of the whole flow equation
in terms of $\tilde\rho$ yields the flows of the different running couplings
$\lambda_n$. 

First, we write down the explicit expression for the Yukawa
coupling, which in the SYM regime reduces to
\begin{equation}
\partial_t h^2=(\eta_\phi+\eta_\mathrm L+\eta_\mathrm R-1)h^2.
\end{equation}
This tells us that an interacting fixed point ($h\neq 0$) can only occur if
\begin{equation}\label{eq:yuksym}
\eta_\phi+\eta_\mathrm L+\eta_\mathrm R=1.
\end{equation}
Conversely, if a fixed point exists in the SYM regime, the sum rule
\eqref{eq:yuksym} has to be satisfied by the anomalous dimensions. This
statement holds exactly in the present truncation but may receive corrections
from higher orders. We comment further on the relevance of this sum rule in
the conclusions.  The expressions for the anomalous dimensions read
\begin{eqnarray}
\eta_\phi&=&\frac{h^2}{3\pi^2}(5-\eta_\mathrm L-\eta_\mathrm R),\\
\eta_\mathrm L&=&\frac{h^2}{6\pi^2}(4-\eta_\mathrm L)\frac{1}{(1+m^2)^2},\\
\eta_\mathrm R&=&\NL  \frac{h^2}{6\pi^2}(4-\eta_\mathrm L)\frac{1}{(1+m^2)^2}.
\end{eqnarray}
This is a linear system of equations which can be solved analytically. Its
solution expresses the anomalous dimensions in terms of the couplings
$h^2$ and~$m^2$,
\begin{eqnarray}
\eta_\phi&=&\frac{2h^2(2h^2(\NL +1)-15\pi^2(1+m^2)^2)}{h^4(\NL
+1)-18\pi^4(1+m^2)^2},\\
\eta_\mathrm L&=&\frac{h^2(5h^2-12\pi^2)}{h^4(\NL +1)-18\pi^4(1+m^2)^2},\\
\eta_\mathrm R&=&\NL \frac{h^2(5h^2-12\pi^2)}{h^4(\NL +1)-18\pi^4(1+m^2)^2}.
\end{eqnarray}
These expressions can be plugged into the sum rule (\ref{eq:yuksym}),
resulting in a conditional fixed point for $h^2$
depending on the size of $m^2$, which reads
\begin{widetext}
\begin{eqnarray}\label{eq:h2cond}
h^2_{\mathrm{cond}}&=&\frac{3\pi^2}{8(\NL +1)} \bigg\{ 7 + 5 m (2 + m) + 2 \NL -\sqrt{
 33 + m (2 + m) (54 + 25 m (2 + m)) + 12 \NL  + 4 m (2 + m) \NL + 4 \NL ^2}
\bigg\}.\nonumber 
\end{eqnarray}
\end{widetext}
For another solution with a positive root, we have not been able to identify a
true fixed point of the full system by numerical means. Hence, this solution
is ignored in the following. The solution
with the negative root, however, does give a fixed point and will be analyzed in
the following. This solution is positive for all $m^2>0$ and monotonously
increasing. It ranges between $h^2(m^2=0)=\frac{3\pi^2}{8}\left(7 +
2\NL - \sqrt{33 + 4 \NL(3 + \NL)}\right)/(\NL+1)$  and
$h^2(m^2=\infty)=\frac{3\pi^2}{5}$, which is very convenient because it
ensures that the fixed point value for $h^2$ is bounded from above and from
below in a very narrow window for all $m^2$.
%
%
A true fixed point of the system requires fixed points for all scalar
couplings $(m^2(=\lambda_1),\lambda_2,\lambda_3,...)$. The flow equation of a
coupling $\lambda_n$ is always a function of the lower order couplings from
the effective potential up to $\lambda_{n+1}$, i.e.\
\begin{equation}
\partial_t \lambda_n = f_n(h^2,\lambda_1,...,\lambda_{n+1}).\label{eq:patlambda}
\end{equation}
Inserting the conditional fixed point for the Yukawa couplings
\eqref{eq:h2cond} into these flows leaves us with the problem of searching for
a scalar fixed-point potential. This problem is familiar from scalar theories
where the Wilson-Fisher fixed point follows from an equation similar to
\Eqref{eq:patlambda} (with $h^2=0$, of course).  We solve the fixed-point
equations $\pat \lambda_n=0$ approximately by a polynomial expansion of the
potential up to some finite order $n\leq n_{\text{max}}$. Dropping the
higher-order couplings $\lambda_{n>n_{\text{max}}}=0$, the resulting system of
fixed-point equations can be solved explicitly. We find suitable fixed-point
solutions in the symmetric regime for $\NL \in \{1,2\}$. The non-universal
fixed-point values as well as the universal values for the anomalous
dimensions at the fixed point and the first two critical exponents can be read
off from Table~\ref{tab:fpvaluesSYM}. For these results, we have expanded the
effective potential up to $\lambda_6$ at next-to-leading order in the
derivative expansion and computed the corresponding stability matrix,
c.f.~\Eqref{eq:lin}, including the Yukawa coupling flow. We would like to
stress that the fixed-point equations in the present case are technically much
more involved in comparison with those of scalar O$(N)$ models in a similar
approximation, as the insertion of the conditional Yukawa coupling fixed point
introduces a much higher degree of nonlinearity.

\begin{table}[ht]
\caption{\label{tab:fpvaluesSYM} Fixed-point values and critical exponents in
the SYM regime. Fixed points corresponding to a second-order phase transition
of the system exist in this regime only for $\NL=1,2$.}
\begin{ruledtabular}\begin{tabular}{ccccccccc}
$\NL $ &$h^2_\ast$ & $m^{ 2}_\ast$ &$\lambda_2^\ast$ & $\eta^*_\phi$ &
$\eta^*_\mathrm L$ & $\eta^*_\mathrm R$ & $\nu$ & $\omega$\\ \hline
$1$ & 4.496& 0.326 & 5.099 & 0.716 & 0.142 & 0.142 & 1.132 & 0.786 \\
$2$ & 3.364& 0.104 & 3.643 & 0.512 & 0.162 & 0.325 & 1.100 & 0.809
\end{tabular}\end{ruledtabular}
\end{table}
We emphasize that a corresponding scalar O($2\NL$) model does not exhibit a
fixed-point potential in the SYM regime but only in the SSB regime. We
conclude that the nature of the phase transition and the corresponding
critical behavior is characteristic for our fermionic model. In particular for
small $\NL$, the fermionic fluctuations contribute with a comparatively large
weight to the critical behavior, as discussed in Fig.~\ref{fig:loops}. 

\begin{figure}[!ht]
 \includegraphics[scale=0.8]{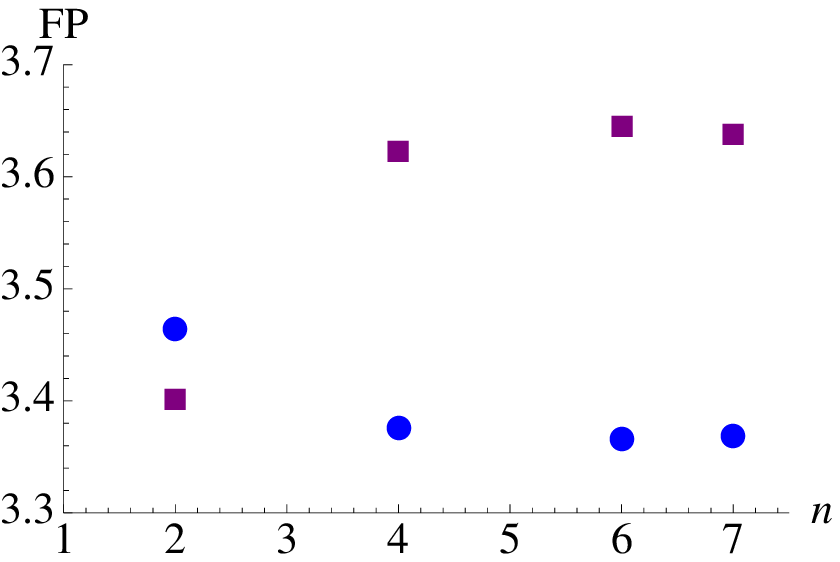}
 \includegraphics[scale=0.8]{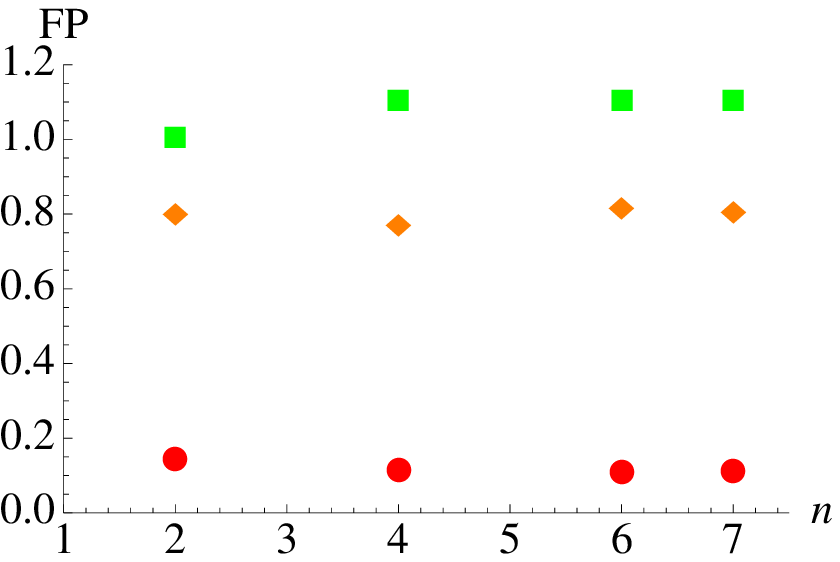}
 \caption{Error estimate for $N_\mathrm{L}=2$: fixed-point values
   $h^2_\ast,m^{ 2}_\ast,\lambda_2^\ast$ and critical exponents $\nu,
   \omega$ as a function of the highest-order term $n$ in the $\rho$ expansion
   of the effective potential.  Left panel: $h_\ast^2$ (blue dots),
   $\lambda_2^\ast$ (purple squares). Right panel: $m^2_\ast$ (red dots),
   $\nu$ (green squares), $\omega$ (orange diamonds).}\label{fig:errest}
\end{figure}

In order to estimate the error on our results arising from the polynomial
expansion of the effective potential, we study the convergence of the
fixed-point values and of the critical exponents as a function of increasing
truncation order for $\NL=2$. Figure \ref{fig:errest} displays our
results for a truncation beyond order $\rho^n\sim \phi^{2n}$ for
$n=2,4,6,7$. All quantities show a satisfactory convergence with a variation
on the 1\%\ level among the highest-order truncations.


Of course, the full truncation error introduced by the derivative expansion is
much harder to determine and depends on the specific quantity. By analogy with
the bosonic O($N$) models, we expect the leading critical exponent to be our
most accurate quantity. On the same truncation level, the critical exponent
$\nu$ in O($N$) models agrees with the best known value already on the
3\%\ level, see \cite{Benitez:2009xg}. The subleading
exponents as well as the anomalous dimensions usually are less well
approximated to this order of the derivative expansion and require more
refined techniques for a better resolution of the momentum dependence, e.g.,
as suggested in \cite{Blaizot:2005xy,Benitez:2009xg}.

\subsection{The regime of spontaneous symmetry breaking (SSB)}
For increasing left-handed fermion number, the scalar fixed-point potential
must eventually lie in the regime of spontaneous symmetry breaking. This is
already obvious from the structure of the flow equations: for large $\NL$, the
Goldstone-like fluctuation modes dominate the flow of the effective potential
\eqref{eq:potflow} (the $\NL$-dependent fermionic contribution in
\Eqref{eq:potflow} is field independent and can be dropped). Therefore, the
potential flow of our model approaches that of an O($2\NL$) model in the limit
$\NL\to\infty$. The latter is known to exhibit a Wilson-Fisher fixed-point
potential in the SSB regime with a nonzero $\kappa^\ast>0$.

\begin{figure}[!ht]
 \includegraphics[scale=0.8]{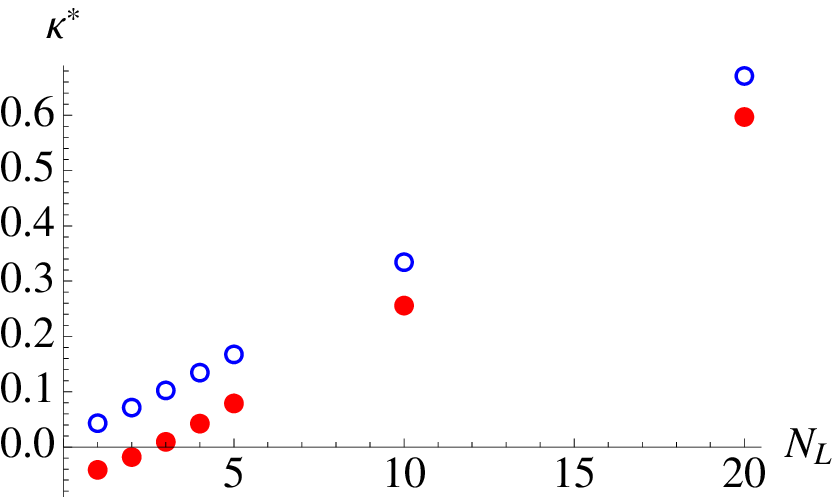}
 \includegraphics[scale=0.8]{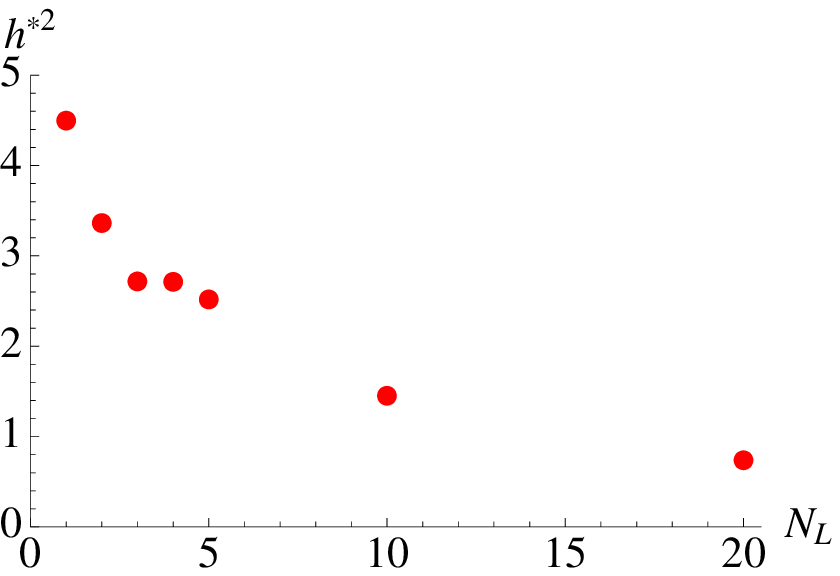}
\caption{\label{fig:nlkappa_nlh2}Nonuniversal fixed point values in the SSB
  regime. Left panel: the potential minimum $\kappa^\ast$ in our U$(\NL
  )_\mathrm{L}\otimes$U$(1)_\mathrm{R}$ model (red dots) are compared with
  those of the O$(2 \NL )$ model (blue circles). The negative $\kappa^\ast$
  values for $\NL<3$ are particular to our model and correspond to the fact
  that the fixed-point potential is in the symmetric regime. This transition
  near $\NL\lesssim 3$ also causes a kink in the $\NL$ dependence of
  $h^\ast{}^2$ (right panel).}
\end{figure}

\begin{figure}[!ht]
 \includegraphics[scale=0.8]{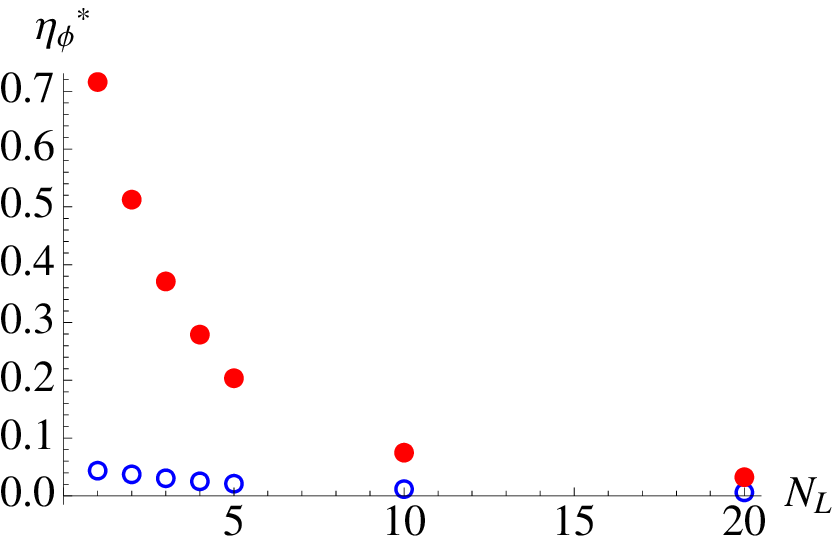}
 \includegraphics[scale=0.8]{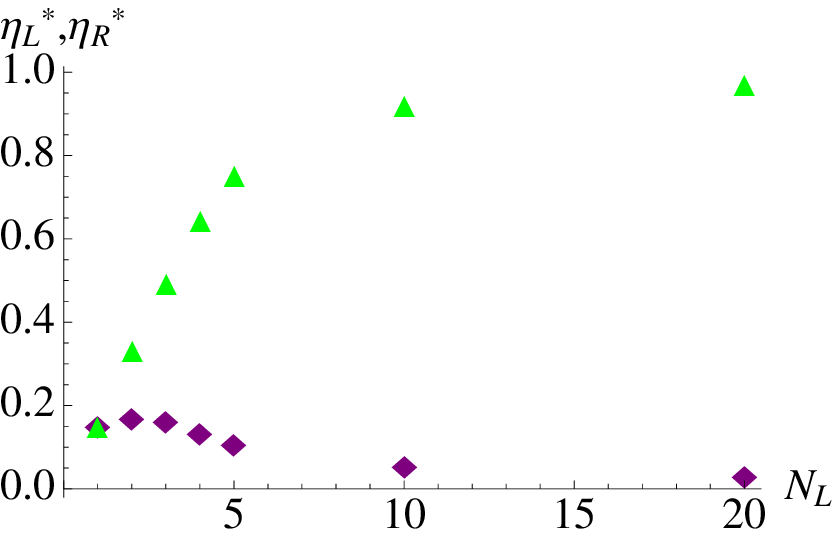}
\caption{\label{fig:nleta}Left: $\eta^*_\phi$ in the U$(\NL
  )_\mathrm{L}\otimes$U$(1)_\mathrm{R}$ model (red dots) in comparison with
  that of the analogous O$(2 \NL )$ model (blue circles). Right:
  $\eta^*_\mathrm L$ (purple diamonds) and $\eta^*_\mathrm R$ (green
  triangles) in the chiral Yukawa model.}
\end{figure}

\begin{figure}[!ht]
\includegraphics[scale=0.8]{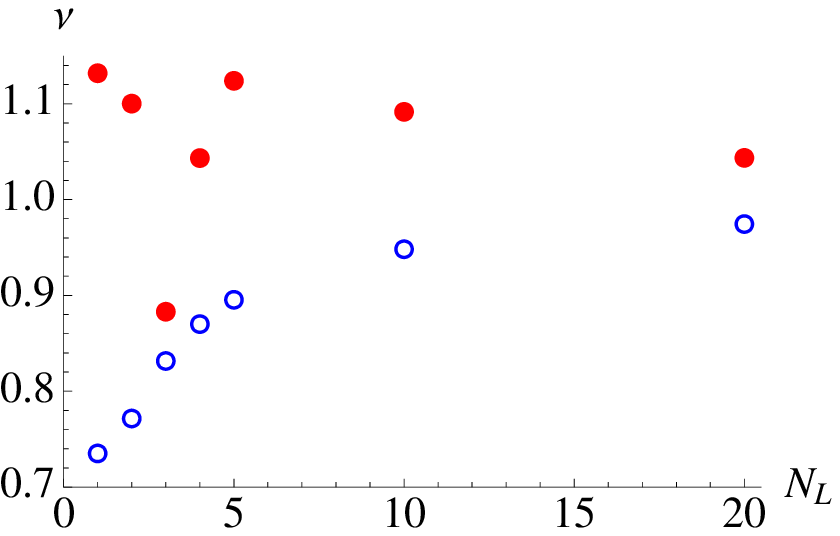}
 \includegraphics[scale=0.8]{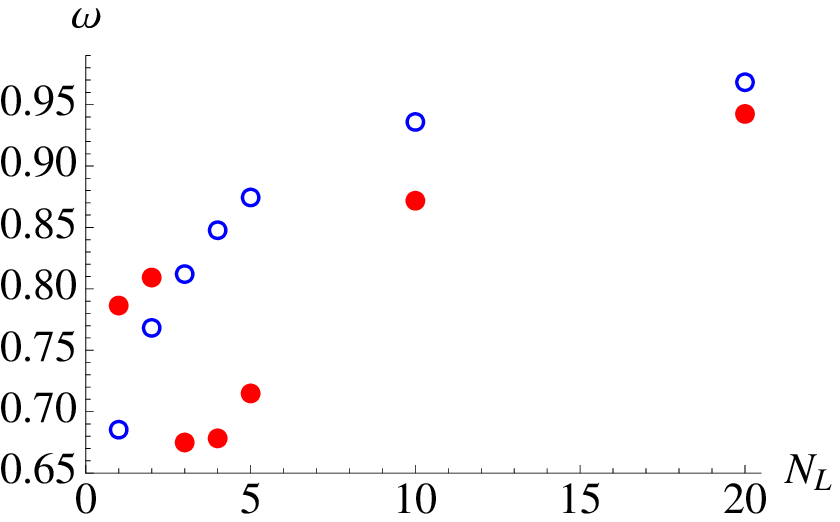}
 \caption{\label{fig:nlnu_nlomega}Critical exponents as a function of $\NL$ in 
   the chiral U$(\NL )_\mathrm{L}\otimes$U$(1)_\mathrm{R}$ model (red dots) in
   comparison with those of the analogous O$(2\NL )$ model (blue circles). The
 rapid change of the exponents occurs near $\NL\simeq3$, where the fixed-point
 potential shows a transition from the symmetric to the SSB regime.}
\end{figure}

We observe the transition of the fixed-point potential from the
symmetric to the SSB regime already near $\NL=3$. The properties of
the Wilson-Fisher fixed point in the analogous bosonic model are, of course,
quantitatively modified by the presence of fermionic fluctuations, but its
basic characteristics are otherwise left intact. Based on these simple
observations,  we continue with an analysis of the fixed-point structure in
the SSB regime in the remainder of this section.
For the SSB phase where the effective potential $u$ is minimal at a nonzero
value $\kappa>0$,
we use the expansion (dropping the subscript $k$ again for simplicity)
\begin{equation}
u=\sum_{n=2}^{N_\text{p}}\frac{\lambda_{n}}{n!}(\tilde{\rho}
-\kappa)^n\label{eq:uexpSSB}=\frac{\lambda_{2}}{2!}(\tilde{\rho}
-\kappa)^2+\frac{\lambda_{3}}{3!}(\tilde{\rho}-\kappa)^3+\dots
\end{equation}
For the flow of $\kappa$, we use the fact that the first derivative of $u$
vanishes at the minimum, $u'(\kappa)=0$. This implies
\begin{eqnarray}
 0= \partial_t u'(\kappa)&=&\partial_t u'(\tilde\rho)|_{\tilde\rho=\kappa}
  +(\partial_t \kappa)u''(\kappa)\nonumber\\
	\Rightarrow \partial_t \kappa&=&-\frac{1}{u''(\kappa)}\partial_t
        u'(\tilde\rho)|_{\tilde\rho=\kappa}\,. \label{eq:kappa}
\end{eqnarray}
The flow equations in the SSB regime are more involved due to additional loop
contributions which arise from the coupling to the vev. This higher degree of
nonlinearity inhibits a simple analytical study of the fixed-point structure
at NLO in the derivative expansion. Instead, we use an iterative method,
starting at the Wilson-Fisher fixed point for the analogous O$(2 \NL )$
model. This fixed point can be obtained in our system by setting the Yukawa
coupling to zero, $h^2=0$.\footnote{This offers also the possibility of a
  cross check: a comparison of our O$(2 \NL )$-model results to the results
  which are given in \cite{Benitez:2009xg} reveals a very satisfactory
  precision. The remaining minor deviations can fully be attributed to the
  fact that we use a simple NLO-derivative expansion truncated at
  $\lambda_6$.}  Starting at this Wilson-Fisher
fixed point of the reduced scalar system, we can obtain a fixed point of the
full chiral Yukawa system by numerical iteration. This confirms that the
presence of the fermions generically shifts the scalar fixed-point values only
slightly. However, our chiral Yukawa system represents a different
universality class, and so the critical exponents and the anomalous dimensions
are special to our system. Numerical results are displayed in
Figs.~\ref{fig:nlkappa_nlh2}-\ref{fig:nlnu_nlomega}.

In Fig.~\ref{fig:nlkappa_nlh2}, the fixed-point values of $\kappa$ and $h^2$
are plotted for different $N_{\mathrm{L}}$. Note that $\kappa^{\ast}$ does not
satisfy the constraint $\kappa\geq 0$ for $N_{\mathrm{L}}<3$; these negative
values are therefore unphysical. Of course, this negative branch corresponds
to the fixed-point scenario in the symmetric regime discussed in the previous
section.  As expected, the fixed-point values of our U$(\NL
)_\mathrm{L}\otimes$U$(1)_\mathrm{R}$ model (red dots) change only slightly in
comparison with the analogous O$(2 \NL )$ model (blue circles). Both models
were investigated with anomalous dimensions and expanded up to order $\sim
\phi^{12}$ in the effective potential. It should be stressed that the
fixed-point values themselves are nonuniversal quantities depending on the
details of the regulator.

By contrast, the fixed-point values of the anomalous dimensions as plotted in
Fig.~\ref{fig:nleta} are universal (even though slight regulator dependencies
can be induced by the truncation).  The left panel shows $\eta_{\phi}^{\ast}$
for our U$(\NL )_\mathrm{L}\otimes$U$(1)_\mathrm{R}$ model (red dots) in
comparison with the analogous O$(2 \NL )$ model (blue circles). Whereas
$\eta_\phi^\ast$ in both models approaches a common value for large $\NL$, we
observe larger differences for smaller $\NL$ which can directly be attributed
to the fermionic loop contributions.  The fermion anomalous dimensions
$\eta_{\mathrm{L}}^{\ast}$ and $\eta_{\mathrm{R}}^{\ast}$ are shown in the
right panel (purple diamonds and green triangles, respectively). For $\NL=1$,
both anomalous dimensions agree as this corresponds to the left/right
symmetric point. For larger $\NL$, $\eta_{\text{R}}^\ast$ becomes
significantly larger, as the massless fermion and boson degrees of freedom
contribute to the loop diagrams with a weight $\sim \NL$. 

The critical exponents $\nu$ and $\omega$ are shown in
Fig.~\ref{fig:nlnu_nlomega}.  Again, the red dots represent the values for our
chiral U$(\NL )_\mathrm{L}\otimes$U$(1)_\mathrm{R}$ model, which is compared
with those of the analogous O$(2 \NL )$ model (blue circles). The results of
both models approach each other for large $\NL$ as expected but show sizable
deviations at smaller $\NL$. We observe a rapid change of the critical
exponents near $\NL\simeq 3$, where the effective fixed point potential
changes from the symmetric to the SSB regime. This strong variation is rather natural,
as the structure of the fixed-point equations varies significantly across this
transition.

\begin{table}[!ht]
\caption{\label{tab:fpvaluesSSB}Fixed point values and critical exponents in
the SSB regime.}
\begin{ruledtabular}\begin{tabular}{ccccccccc}
$\NL $ &$h^2_\ast$ & $\kappa_\ast$ &$\lambda_2^\ast$ & $\eta^*_\phi$ &
$\eta^*_\mathrm L$ & $\eta^*_\mathrm R$ & $\nu$ & $\omega$\\ \hline
$3$ & 2.718 & 0.009 & 2.967 & 0.371 & 0.154 & 0.487 & 0.883 & 0.675 \\
$4$ & 2.713 & 0.042 & 2.954 & 0.279 & 0.125 & 0.637 & 1.043 & 0.678 \\
$5$ & 2.519 & 0.079 & 2.717 & 0.204 & 0.100 & 0.746 & 1.124 & 0.715 \\
$10$ & 1.452 & 0.256 & 1.506 & 0.075 & 0.046 & 0.913 & 1.092 &0.872 \\
$20$ & 0.739 & 0.597 & 0.752 & 0.032 & 0.022 & 0.963 & 1.043 & 0.942 \\
$50$ & 0.296 & 1.612 & 0.298 & 0.012 & 0.009 & 0.986 & 1.017 & 0.978 \\
$100$ & 0.148 & 3.301 & 0.149 & 0.006 & 0.004 & 0.993 & 1.008 & 0.989
\end{tabular}\end{ruledtabular}
\end{table}

We believe that especially the results for the universal fixed-point anomalous
dimensions and critical exponents as summarized in Tab.~\ref{tab:fpvaluesSSB}
can provide for first benchmarks for these new universality classes. For any
other nonperturbative tool for chiral fermions near the symmetry-breaking
phase transition, these universality classes can serve as a useful testing
ground. As in the symmetric regime discussed above, we consider our results for
the exponent $\nu$ to be reliable on the few-percent level, whereas the
subleading exponent as well as the anomalous dimensions might be somewhat less
accurate.

\section{Conclusions}
\label{sec:conclusions}

We have investigated the critical behavior of three-dimensional relativistic
fermion models with a U($\NL$)$_\mathrm{L} \times \mathrm U(1)_{\mathrm{R}}$
chiral symmetry. We have designed this class of models to exhibit similarities
to the Higgs-Yukawa sector of the standard model of particle physics. As a
fundament of this model building, we have classified all possible four-fermion
interaction terms invariant under this chiral symmetry, and also determined
the corresponding discrete symmetries. We have identified a scalar
parity-conserving channel similar to the standard-model Higgs scalar.  For
sufficiently strong correlations in this channel, a second-order phase
transition into the chiral-symmetry broken phase can occur which is a 3d
analog of the electroweak phase transition. Using the functional RG, we
have computed the critical behavior of this phase transition in terms of the
critical correlation-length exponent $\nu$, the subleading exponent $\omega$,
and the fermion and scalar anomalous dimensions as a function of $\NL\geq1$.

Whereas the standard model is defined with a fundamental Higgs scalar, we have
here started with only fundamental fermion degrees of freedom. The resulting
Higgs field arises as a scalar bi-fermionic composite upon strong fermionic
correlations. Therefore, the relation between our original fermion model and
the resulting Yukawa system is similar to that between top-quark condensation
models \cite{BHL} and the standard-model Higgs sector. From a more general RG
perspective, however, the purely fermionic models can anyway be viewed as just
a special case of the more general Yukawa models
\cite{ZinnJustin:1991yn,Hasenfratz:1991it} supplemented with nonuniversal
compositeness conditions. As long as the compositeness scale in the deep UV
remains unresolved, there is no real difference between the purely fermionic
or the Yukawa-model language.

For our quantitative results, we have used a consistent and systematic
expansion scheme of the effective action in terms of a nonperturbative
derivative expansion. Whereas there is an extensive body of circumstantial
evidence in the literature that this expansion is suitably adjusted to the
relevant degrees of freedom of Yukawa systems, a practical test for
convergence is problematic in the present case. This is because the
leading-order of the expansion (defined by setting all
$\eta_{\phi,\text{L,R}}=0$) does not support the desired fixed point. The
latter becomes visible only from next-to-leading order on, due to the
structure of the Yukawa flow. Hence, a straightforward convergence test in
principle requires a NNLO calculation. Instead, the reliability of the results
can also be verified indirectly: first of all, the derivative expansion is
based on the implicit assumption that momentum dependencies of operators do
not grow large. This includes the kinetic terms, such that self-consistency of
the derivative expansion requires that the anomalous dimensions satisfy
$\eta_{\phi,\text{L,R}}\lesssim 1$, as is the case in our
calculations. Second, our models can always be compared to the purely bosonic
limit of scalar O($N$) models where the quantitative reliability of the
derivative expansion has been verified to a high level of significance. Third,
we observe very good convergence properties of the polynomial expansion of the
effective potential which is again a strong signature of self-consistency.

From our classification of all fermionic interaction terms compatible with the
required symmetries, it is clear that the Higgs-like condensation channel is
not the only possible channel. Aside from vector-like channels, there are two
further pseudo-scalar channels, cf. \Eqref{eq:pseudoscalar}, and further
scalar and pseudo-scalar channels in the flavor-singlet Fierz transforms of
Eqs.~\eqref{eq:scalar}-\eqref{eq:pseudovector}. In fact, the present analysis
is a restricted study of a particular condensation process. We expect that the
phase diagram of the general model is much more involved and might exhibit a
variety of possible phases and corresponding transitions. This phase diagram
is parameterized by up to 6 independent couplings being associated with the
linearly independent fermionic interactions.  The calculation of the true
condensation channel for a given set of initial couplings remains a
challenging problem. As such a problem of competing order parameters is well
known also in other systems, e.g., in the Hubbard model
\cite{Halboth:2000zza,Salmhofer:2001tr,Krahl:2009}, the present system can serve as a
rich and controllable model system.

A special feature of our model arises in the symmetric regime: here, a fixed
point within our truncation implies a sum rule for the anomalous dimensions,
$\eta_\phi + \eta_{\text{L}}+ \eta_{\text{R}}=1\equiv 4-d$. This sum rule is
relevant, since the underlying balancing between anomalous dimensions and
dimensional power-counting scaling can be a decisive feature of many other
models as well. Most prominently, the asymptotic-safety scenario in quantum
gravity \cite{Reuter:1996cp} as well as in extra-dimensional Yang-Mills
theories \cite{Gies:2003ic} requires similar sum rules to be satisfied. In
contrast to these latter models, the present models for $\NL=1,2$ can serve as
a much simpler example for a test of this sum rule at a fixed point. A
verification of this sum rule also by other nonperturbative tools can shed
light on this important mechanism to generate RG fixed points.

This is another reason why we believe that the new universality classes
defined by our models can serve as prototypes for studies of strongly
correlated chiral fermions in general and of nonperturbative features of
standard-model-like chiral symmetry breaking in particular.

\acknowledgments


For their hospitality, M.M.S.~thanks Roberto Percacci and SISSA where part of
this work was completed.  This work was supported by the DFG under contract
No. Gi 328/5-1 (Heisenberg program), FOR 723, and GK1523/1.

\appendix
\section{Flow equations and anomalous dimensions}\label{sec:flowequations}
The derivation of the flow of the dimensionless effective potential
$\partial_tu_k$ \eqref{eq:potflow} and of the anomalous dimensions
$\eta_{\mathrm{L}}$ and $\eta_{\mathrm{R}}$ \eqref{eq:anomalousDimensions}
are explained in the appendix of \cite{Gies:2009sv}.
Consequently the flow of the coupling constants $\lambda_i$ and
the flow of the squared mass $m^2$ in the symmetric regime and the flow of the
dimensionless squared vacuum expectation value $\kappa$ in the SSB regime
are the same as in \cite{Gies:2009sv}.

For deriving the flow of the squared Yukawa coupling constant $h^2$ we split
the bosonic field into its vev $v$ and the deviation from the vev (the relation
between the dimensionful vev $v$ and the dimensionless squared vev $\kappa$ is
$\kappa = \frac{1}{2} Z_{\phi} k^{2-d} v^2$),
\begin{widetext}
\begin{equation}\label{eq:decomposition}
\phi(p)=\frac{1}{\sqrt{2}}
\begin{pmatrix}
\phi_1^1(p)+\ir \phi_2^1(p)\\
\phi_1^2(p)+\ir \phi_2^2(p)\\
\vdots\\
\phi_1^{N_\mathrm L}(p)+\ir \phi_2^{N_\mathrm L}(p)
\end{pmatrix}=\frac{1}{\sqrt{2}}
\begin{pmatrix}
v\\
0\\
\vdots\\
0
\end{pmatrix}\delta(p)+\frac{1}{\sqrt{2}}
\begin{pmatrix}
\Delta\phi_1^1(p)+\ir \Delta\phi_2^1(p)\\
\Delta\phi_1^2(p)+\ir \Delta\phi_2^2(p)\\
\vdots\\
\Delta\phi_1^{N_\mathrm L}(p)+\ir \Delta\phi_2^{N_\mathrm L}(p)
\end{pmatrix}.
\end{equation}
We are mainly interested in the Yukawa coupling between the fermions and the
Goldstone boson. Thus we use the $\Delta\phi_2^1$ part for the projection. Using
the truncation \eqref{eq:SULtruncation} and comparing with
the Wetterich equation we get
\begin{equation}\label{eq:yukawaFlow}
\partial_t\bar{h}_k=-\frac{\ir}{2}\left.\frac{\overrightarrow{\delta}}{\delta\bar{
\psi}_{\mathrm{L}}^1(p)}\frac{\sqrt{2}\overrightarrow{\delta}}{
\delta\Delta\phi_2^1(p')}
\mathrm{STr}\left[\tilde{\partial}_t\ln(\Gamma_k^{(2)}+R_k)\right]
\frac{\overleftarrow{\delta}}{\delta\psi_{\mathrm{R}}(q)}\right|_{\overset{\psi_
{\mathrm{R}}^a=\psi_{\mathrm{L}}^a=\Delta\phi=0}{p'=p=q=0}}.
\end{equation}
Next, we split $(\Gamma_k^{(2)}+R_k)$ into a propagator part $\mathcal{P}$, which
contains only the vev, and a fluctuation part $\mathcal{F}$, which contains the
fluctuating fields. Inserting this into Eq.~\eqref{eq:yukawaFlow} the
expansion of the logarithm reads
\begin{equation}\label{eq:logarithmExpand}
\ln\left(\Gamma_k^{(2)}+R_k\right)=\ln\left[ \mathcal{P}\left(
1+\frac{\mathcal{F}}{\mathcal{P}} \right)
\right]=\ln(\mathcal{P})+\frac{\mathcal{F}}{\mathcal{P}}-\frac{1}{2}\left(
\frac{\mathcal{F}}{\mathcal{P}} \right)^2+\frac{1}{3}\left(
\frac{\mathcal{F}}{\mathcal{P}} \right)^3-\ldots
\end{equation}
Only the term to third power survives the projection. Performing the matrix
calculations and taking the supertrace, we get
\begin{equation*}
\partial_t\bar{h}_k^2=\int
\frac{d^dp}{(2\pi)^d}\tilde{\partial}_t\frac{h_k^4U_{k}''v^2}{\left(Z_{\phi}P_{
\mathrm{B}}(p)+U_{k}'+U_{k}''v^2\right)\left(Z_{\phi}P_{\mathrm{B}}(p)+U_{k}
'\right)\left(Z_ { \mathrm { L
}}Z_{\mathrm{R}}P_{\mathrm{F}}(p)+\frac{\bar{h}_k^2}{2}v^2\right)}.
\end{equation*}
The potential on the right-hand side is evaluated at the minimum
$\frac{1}{2}v^2$. Using the threshold function and switching over to
dimensionless quantities, we end up with Eq.~\eqref{eq:hquadratFlow}.

For the derivation of the flow of $\eta_{\phi}$, we use the decomposition
of \eqref{eq:decomposition}. Again we use $\Delta\phi_2^1$ for the projection
and expand the logarithm as in \eqref{eq:logarithmExpand}. This time only
the quadratic term survives the projection,
\begin{equation*}
\left. \partial_tZ_{\phi}=-\frac{1}{4}\frac{\partial}{\partial
  p^2}\frac{\delta}{\delta \Delta \phi_2^1(p)}\frac{\delta}{\delta \Delta
  \phi_2^1(q)}\mathrm{STr}\left[ \tilde{\partial}_t\left(
\frac{\mathcal F}{\mathcal P}
\right)^2
  \right]\right|_{\psi=\Delta \phi =0=p=q}.
\end{equation*}
\end{widetext}
After performing the matrix calculations and taking the supertrace, we use
$\eta_{\phi}=-\frac{\partial_tZ_{\phi}}{Z_{\phi}}$ and switch over to
dimensionless quantities. By use of the threshold functions
\eqref{eq:etaThresholds}, we obtain Eqs.~\eqref{eq:anomalousDimensions} in the
main text.


\begin{thebibliography}{12}
\expandafter\ifx\csname natexlab\endcsname\relax\def\natexlab#1{#1}\fi
\expandafter\ifx\csname bibnamefont\endcsname\relax
  \def\bibnamefont#1{#1}\fi
\expandafter\ifx\csname bibfnamefont\endcsname\relax
  \def\bibfnamefont#1{#1}\fi
\expandafter\ifx\csname citenamefont\endcsname\relax
  \def\citenamefont#1{#1}\fi
\expandafter\ifx\csname url\endcsname\relax
  \def\url#1{\texttt{#1}}\fi
\expandafter\ifx\csname urlprefix\endcsname\relax\def\urlprefix{URL }\fi
\providecommand{\bibinfo}[2]{#2}
\providecommand{\eprint}[2][]{\url{#2}}

\bibitem{Stueckelberg:1953dz}
  E.~C.~G.~Stueckelberg and A.~Petermann,
  Helv.\ Phys.\ Acta {\bf 26}, 499 (1953).

\bibitem{Gell-Mann:fq}
M.~Gell-Mann and F.~E.~Low,
Phys.\ Rev.\  {\bf 95}, 1300 (1954).

\bibitem{Bogolyubov:1956gh}
  N.~N.~Bogolyubov and D.~V.~Shirkov,
  Nuovo Cim.\  {\bf 3}, 845 (1956).

\bibitem{Gies:2009hq}
  H.~Gies and M.~M.~Scherer,
  arXiv:0901.2459 [hep-th].
\bibitem{Gies:2009sv}
  H.~Gies, S.~Rechenberger and M.~M.~Scherer,
  arXiv:0907.0327 [hep-th],
  M.~M.~Scherer, H.~Gies and S.~Rechenberger,
  arXiv:0910.0395 [hep-th].


\bibitem{Rosa:2000ju}
  L.~Rosa, P.~Vitale and C.~Wetterich,
  Phys.\ Rev.\ Lett.\  {\bf 86}, 958 (2001)
  [arXiv:hep-th/0007093]; 
%
  F.~Hofling, C.~Nowak and C.~Wetterich,
  Phys.\ Rev.\  B {\bf 66}, 205111 (2002)
  [arXiv:cond-mat/0203588].


\bibitem{HandsEtAl}
  S.~Hands, A.~Kocic and J.~B.~Kogut,
  Annals Phys.\  {\bf 224}, 29 (1993)
  [arXiv:hep-lat/9208022];
  K.~I.~Aoki, K.~i.~Morikawa, J.~I.~Sumi, H.~Terao and M.~Tomoyose,
  Prog.\ Theor.\ Phys.\  {\bf 97}, 479 (1997)
  [arXiv:hep-ph/9612459];
  J.~A.~Gracey,
  Int.\ J.\ Mod.\ Phys.\  A {\bf 9}, 567 (1994)
  [arXiv:hep-th/9306106];
  L.~Karkkainen, R.~Lacaze, P.~Lacock and B.~Petersson,
  Nucl.\ Phys.\  B {\bf 415}, 781 (1994)
  [Erratum-ibid.\  B {\bf 438}, 650 (1995)]
  [arXiv:hep-lat/9310020];
  J.~A.~Gracey,
  Int.\ J.\ Mod.\ Phys.\  A {\bf 9}, 727 (1994)
  [arXiv:hep-th/9306107];
  A.~N.~Vasiliev, S.~E.~Derkachov, N.~A.~Kivel and A.~S.~Stepanenko,
  Theor.\ Math.\ Phys.\  {\bf 94}, 127 (1993)
  [Teor.\ Mat.\ Fiz.\  {\bf 94}, 179 (1993)].


\bibitem{Pisarski:1984dj}
  R.~D.~Pisarski,
  Phys.\ Rev.\  D {\bf 29}, 2423 (1984).
\bibitem{Appelquist:1985vf}
  T.~Appelquist, M.~J.~Bowick, E.~Cohler and L.~C.~R.~Wijewardhana,
  Phys.\ Rev.\ Lett.\  {\bf 55}, 1715 (1985); 
  T.~W.~Appelquist, M.~J.~Bowick, D.~Karabali and L.~C.~R.~Wijewardhana,
  Phys.\ Rev.\  D {\bf 33}, 3704 (1986); 
  Phys.\ Rev.\  D {\bf 33}, 3774 (1986); 
  T.~Appelquist, D.~Nash and L.~C.~R.~Wijewardhana,
  Phys.\ Rev.\ Lett.\  {\bf 60}, 2575 (1988); 
  D.~Nash,
  Phys.\ Rev.\ Lett.\  {\bf 62}, 3024 (1989); 
  G.~W.~Semenoff and L.~C.~R.~Wijewardhana,
  Phys.\ Rev.\ Lett.\  {\bf 63}, 2633 (1989).

\bibitem{Gomes:1990ed}
  M.~Gomes, R.~S.~Mendes, R.~F.~Ribeiro and A.~J.~da Silva,
  Phys.\ Rev.\  D {\bf 43}, 3516 (1991); 
  D.~K.~Hong and S.~H.~Park,
  Phys.\ Rev.\  D {\bf 49}, 5507 (1994)
  [arXiv:hep-th/9307186].

\bibitem{Hands:2002dv}
  S.~J.~Hands, J.~B.~Kogut and C.~G.~Strouthos,
  Nucl.\ Phys.\  B {\bf 645}, 321 (2002)
  [arXiv:hep-lat/0208030]; 
  S.~J.~Hands, J.~B.~Kogut, L.~Scorzato and C.~G.~Strouthos,
  Phys.\ Rev.\  B {\bf 70}, 104501 (2004)
  [arXiv:hep-lat/0404013]; 
  C.~Strouthos and J.~B.~Kogut,
  PoS {\bf LAT2007}, 278 (2007)
  [arXiv:0804.0300 [hep-lat]]; 
  S.~Christofi, S.~Hands and C.~Strouthos,
  Phys.\ Rev.\  D {\bf 75}, 101701 (2007)
  [arXiv:hep-lat/0701016].


\bibitem{Maris:1996zg}
  P.~Maris,
  Phys.\ Rev.\  D {\bf 54}, 4049 (1996)
  [arXiv:hep-ph/9606214]; 
  C.~S.~Fischer, R.~Alkofer, T.~Dahm and P.~Maris,
  Phys.\ Rev.\  D {\bf 70}, 073007 (2004)
  [arXiv:hep-ph/0407104]; 
  A.~Bashir, A.~Raya, S.~Sanchez-Madrigal and C.~D.~Roberts,
  arXiv:0905.1337 [hep-ph].

\bibitem{Mavromatos:2003ss}
  N.~E.~Mavromatos and J.~Papavassiliou,
  arXiv:cond-mat/0311421.

\bibitem{Herbut:2002yq}
  I.~F.~Herbut,
  Phys.\ Rev.\  B {\bf 66}, 094504 (2002)
  [arXiv:cond-mat/0202491]; 
  K.~Kaveh and I.~F.~Herbut,
  Phys.\ Rev.\  B {\bf 71}, 184519 (2005)
  [arXiv:cond-mat/0411594].

\bibitem{Herbut:2009qb}
  I.~F.~Herbut, V.~Juricic and B.~Roy,
  Phys.\ Rev.\  B {\bf 79}, 085116 (2009)
  [arXiv:0811.0610 [cond-mat.str-el]]; 
  V.~P.~Gusynin, S.~G.~Sharapov and J.~P.~Carbotte,
  Int.\ J.\ Mod.\ Phys.\  B {\bf 21}, 4611 (2007)
  [arXiv:0706.3016 [cond-mat.mes-hall]].

\bibitem{Wetterich:1993yh}
  C.~Wetterich,
 Phys.\ Lett.\ B {\bf 301}, 90 (1993).


\bibitem{Berges:2000ew}
  J.~Berges, N.~Tetradis and C.~Wetterich,
  Phys.\ Rept.\  {\bf 363}, 223 (2002)
  [arXiv:hep-ph/0005122]; 
  K.~Aoki,
  Int.\ J.\ Mod.\ Phys.\  B {\bf 14}, 1249 (2000); 
  J.~Polonyi,
  Central Eur.\ J.\ Phys.\  {\bf 1}, 1 (2003)
  [arXiv:hep-th/0110026].
  J.~M.~Pawlowski,
  Annals Phys.\  {\bf 322}, 2831 (2007)
  [arXiv:hep-th/0512261]; 
  H.~Gies,
  arXiv:hep-ph/0611146; 
  B.~Delamotte,
  arXiv:cond-mat/0702365; 
H.~Sonoda,
arXiv:0710.1662 [hep-th] (2007).

\bibitem{Litim:2002cf}
  D.~F.~Litim,
  Nucl.\ Phys.\  B {\bf 631}, 128 (2002)
  [arXiv:hep-th/0203006].
\bibitem{Benitez:2009xg}
  F.~Benitez, J.~P.~Blaizot, H.~Chate, B.~Delamotte, R.~Mendez-Galain and
  N.~Wschebor,
  arXiv:0901.0128 [cond-mat.stat-mech].


\bibitem{Jungnickel:1995fp}
  D.~U.~Jungnickel and C.~Wetterich,
  Phys.\ Rev.\  D {\bf 53}, 5142 (1996)
  [arXiv:hep-ph/9505267]; 
  B.~J.~Schaefer and H.~J.~Pirner,
  Nucl.\ Phys.\  A {\bf 660}, 439 (1999)
  [arXiv:nucl-th/9903003]; 
  B.~J.~Schaefer and J.~Wambach,
  Nucl.\ Phys.\  A {\bf 757}, 479 (2005)
  [arXiv:nucl-th/0403039]; 
%
%
  H.~Gies and C.~Wetterich,
  Phys.\ Rev.\  D {\bf 65}, 065001 (2002)
  [arXiv:hep-th/0107221]; 
  Phys.\ Rev.\  D {\bf 69}, 025001 (2004)
  [arXiv:hep-th/0209183]; 
  J.~Braun,
  arXiv:0810.1727 [hep-ph];
J. Braun, arXiv:0908.1543

\bibitem{Braun:2009si}
  J.~Braun,
  arXiv:0908.1543 [hep-ph].


\bibitem{Birse:2004ha}
  M.~C.~Birse, B.~Krippa, J.~A.~McGovern and N.~R.~Walet,
  Phys.\ Lett.\  B {\bf 605}, 287 (2005)
  [arXiv:hep-ph/0406249]; 
  S.~Diehl, H.~Gies, J.~M.~Pawlowski and C.~Wetterich,
  Phys.\ Rev.\  A {\bf 76}, 053627 (2007)
  [arXiv:cond-mat/0703366]; 
%
  Phys.\ Rev.\ A {\bf 76 }, 21602({\em Rap. Comm.}) (2007) 
 [arXiv:cond-mat/0701198]; 
  S.~Floerchinger, M.~Scherer, S.~Diehl and C.~Wetterich,
  arXiv:0808.0150 [cond-mat.supr-con].

\bibitem{Litim:2001up}
  D.~F.~Litim,
  Phys.\ Rev.\  D {\bf 64}, 105007 (2001)
  [arXiv:hep-th/0103195].

\bibitem{Blaizot:2005xy}
  J.~P.~Blaizot, R.~Mendez Galain and N.~Wschebor,
  Phys.\ Lett.\  B {\bf 632}, 571 (2006)
  [arXiv:hep-th/0503103].

\bibitem{BHL}
Y.~Nambu,
 In *Kazimierz 1988, Proceedings, New theories in physics* 1-10; 
V.~A.~Miransky, M.~Tanabashi and K.~Yamawaki,
Phys.\ Lett.\ B {\bf 221}, 177 (1989);
Mod.\ Phys.\ Lett.\ A {\bf 4}, 1043 (1989); 
W.~A.~Bardeen, C.~T.~Hill and M.~Lindner,
Phys.\ Rev.\ D {\bf 41}, 1647 (1990).


\bibitem{ZinnJustin:1991yn}
  J.~Zinn-Justin,
  Nucl.\ Phys.\  B {\bf 367}, 105 (1991).

\bibitem{Hasenfratz:1991it}
  A.~Hasenfratz, P.~Hasenfratz, K.~Jansen, J.~Kuti and Y.~Shen,
  Nucl.\ Phys.\  B {\bf 365}, 79 (1991).

\bibitem{Halboth:2000zza}
  C.~J.~Halboth and W.~Metzner,
  Phys.\ Rev.\ Lett.\  {\bf 85}, 5162 (2000); 
  Phys.\ Rev.\  B {\bf 61}, 7364 (2000).

\bibitem{Salmhofer:2001tr}
  M.~Salmhofer and C.~Honerkamp,
  Prog.\ Theor.\ Phys.\  {\bf 105}, 1 (2001); 
 C.~Honerkamp and M.~Salmhofer, 
Phys. Rev. Lett. 87, 187004 (2001).

\bibitem{Krahl:2009}
H.~C.~Krahl, J.~A.~Müller, C.~Wetterich,
Phys.\ Rev.\ B {\bf 79}, 094526 (2009) 

\bibitem{Reuter:1996cp}
  M.~Reuter,
  Phys.\ Rev.\  D {\bf 57}, 971 (1998)
  [arXiv:hep-th/9605030]; 
O.~Lauscher and M.~Reuter,
Phys.\ Rev.\ D {\bf 65}, 025013 (2002)
[arXiv:hep-th/0108040]; 
%
W.~Souma,
Prog.\ Theor.\ Phys.\  {\bf 102}, 181 (1999)
[arXiv:hep-th/9907027]; 
%
  P.~Forgacs and M.~Niedermaier,
  arXiv:hep-th/0207028; 
  R.~Percacci and D.~Perini,
  Phys.\ Rev.\  D {\bf 68}, 044018 (2003)
  [arXiv:hep-th/0304222]; 
  A.~Codello, R.~Percacci and C.~Rahmede,
  Int.\ J.\ Mod.\ Phys.\  A {\bf 23}, 143 (2008)
  [arXiv:0705.1769 [hep-th]]; 
  D.~Benedetti, P.~F.~Machado and F.~Saueressig,
  arXiv:0901.2984 [hep-th];
  D.~Benedetti, P.~F.~Machado and F.~Saueressig,
  arXiv:0902.4630 [hep-th]; 
  A.~Eichhorn, H.~Gies and M.~M.~Scherer,
  arXiv:0907.1828 [hep-th].


\bibitem{Gies:2003ic}
H.~Gies,
Phys.\ Rev.\ D {\bf 68}, 085015 (2003)
[arXiv:hep-th/0305208].


\end{thebibliography}
\end{document}